Modelling of thermal stratification at the top of a planetary core:

Application to the cores of Earth and Mercury and the thermal coupling with their mantles

J. S. Knibbe[1,2*] and T. Van Hoolst[1,2]

[1] Royal Observatory of Belgium, Brussels, Belgium.

[2] KU Leuven, Leuven, Belgium.

[*]Corresponding author: j.s.knibbe@vu.nl

**Abstract**

We present a new numerical scheme for solving one-dimensional conduction problems and in the radial direction of a spherically symmetric body or shell with variable spatial domain. This numerical scheme adopts a solution of the conduction equation in each interval of the chosen discretization, that is valid if the fluxes at the interval boundaries are constant in time. This 'piece-wise steady flux' (PWSF) numerical scheme is continuous and differentiable in the space domain. These smoothness properties are convenient for implementing the numerical scheme in an energy-conserved thermal evolution approach for the cores of Earth and Mercury, in which a conductive thermally stratified layer is considered that develops below the core-mantle boundary when the heat flux drops below the adiabatic heat flux. The influence of a time-variable thermally stratified region on the general evolution of the planetary body is examined, in comparison to imposing an adiabatic temperature profile for the entire core. Also, the dependence of the model's accuracy to the applied grid size of the numerical scheme is studied. We show that the new numerical model is numerically very efficient.

By considering thermal stratification in a planetary core where the heat flux is lower than the adiabatic heat flux, radial variations in the cooling rate are accounted for whereas otherwise the distribution of energy in the core is fixed by the imposed adiabat. During the growth of the thermally stratified region, the deep part of the core cools more rapidly than the outer part of the core. Therefore, the inner core grows to a larger size and the temperature and heat flux at the core-mantle boundary are higher and larger, respectively, if a thermally stratified region is considered. For the Earth, the implications are likely very minor and can be neglected in thermal evolution studies that are not specifically interested in the thermally stratified region itself. For Mercury, however, these implications are much larger. For example, the age of the inner core can be underestimated by several billion years if thermal stratification is neglected. The consideration of thermal stratification in Mercury's core is also important for the evolution of Mercury's mantle. It increases the mantle temperature, leads to a higher

Rayleigh number and therefore a larger heat flux into the lithosphere and prolongation of mantle convection.

**Highlights**

- We present an efficient numerical scheme for 1-D conduction, smooth in the space domain.
- We study implications of upper core thermal stratification in Earth and Mercury.
- Thermal stratification in Earth's core is unimportant for the planet's evolution.
- Thermal stratification in Mercury's core is important for the planet's evolution.

## 1 Introduction

For both the Earth and Mercury, it has been suggested that the heat flux from the core to the mantle is smaller than the heat flux conducted along the adiabat at the top of the core, such that an upper region of the liquid outer core is thermally stratified.

A stratified upper region in the Earth's liquid outer core is supported by the observation of fluctuations in the geomagnetic field with a period of 60 years, which are likely due to Magnetic-Archimedes-Coriolis (MAC) waves in a stable top layer of the core (e.g. Buffet, 2014). These waves may also be the cause of decadal length-of-day variations, adding further evidence in favour of the existence of a stable layer (Buffet et al., 2016). Heat flux estimates, which range between 5 TW and 17 TW for the core-mantle boundary heat flux and between 9 TW and 15 TW for the adiabatic heat flux at the core-mantle boundary (Jaupart et al., 2015), cannot prove that an upper region of Earth's core is thermally stratified (Labrosse, 2015) but also do not rule out the existence of such a thermally stratified layer (Davies et al., 2015; Gubbins et al., 2015). Because thermal conductivity increases with pressure, it has been suggested that, when volumetrically heterogenous sources of buoyancy are absent in the core, the heat flux would first drop below the adiabatic heat flux at the center of the core (Gomi et al., 2013; Labrosse, 2015). Thermal stratification at the center of Earth's core would make the operation of a dynamo impossible, if volumetrically heterogenous sources of buoyancy are absent (Labrosse, 2015). Based on paleomagnetic evidence of an active geodynamo over the last ~3.5 billion years (Tarduno et al., 2010), Labrosse (2015) argues that it is unlikely that Earth's liquid core experienced thermal stratification before the inner core has started to solidify. The release of energy by the growth of an inner core leads to an increase of thermal buoyancy with depth in the liquid core. Therefore, during the growth of an inner core, thermal stratification may initiate and persist in an upper region of the core while a geodynamo operates in the rest of the liquid core (Davies et al., 2015; Gubbins et al., 2015). Based on the estimates for the growth rate of the inner core and its present-day radius of 1221 km, it is generally accepted that the age of the inner core is much younger than 3.5 billion years (e.g., Labrosse et al., 2001; Labrosse, 2015). Therefore, it is expected that thermal stratification in the upper part of Earth's liquid core is also a relatively recent phenomenon, if it occurs (Labrosse, 2015). The occurrence of thermal stratification in exclusively an upper region of the liquid core does not only depend on the heat flux at the core-mantle boundary, but also on the precise radial profiles of thermal conductivity and thermal expansion that are both prone to debate (Pozzo et al, 2012, 2014; Seagle et al., 2013; Gomi et al., 2013; Labrosse, 2015; Gubbins et al., 2015). Analyses of seismic data suggests that P-waves are slower in an upper layer in the core (e.g. Helffrich and Kaneshima, 2010). This could indicate that this upper region of the core is compositionally different from the rest of Earth's liquid core. It has been suggested that an upper region of the Earth's liquid core is enriched in light elements and compositionally stratified. The possibility of compositional stratification in an uppermost region of Earth's liquid core is, however, also still debated. For example, Buffet and Seagle (2010) and Nagakawa (2018) show that the enrichment of light elements is expected to increase rather than decrease the P-wave velocity of the iron-rich liquid core metal.

Thermal evolution studies of Mercury consistently show that the heat flux through its core-mantle boundary drops below the adiabatic heat flux at the top of the core within the first billion years after Mercury's core-mantle differentiation (e.g., Stevenson et al., 1983; Grott et al., 2011; Michel et al., 2013; Tosi et al., 2013; Knibbe and van Westrenen, 2018). An upper layer that is stable against convection in the upper part of Mercury's liquid core is consistent with the deep-dynamo explanation for Mercury's present-day broad-scale and low-intensity magnetic field (e.g., Christensen, 2006; Christensen and Wicht, 2008; Manglik et al., 2010; Cao et al., 2014; Tian et al., 2015; Takahashi et al., 2019). Additional to thermal stratification, compositional stratification has been suggested to be generated by an iron-snow process that can exist just below Mercury's core-mantle boundary (e.g., Chen et al., 2008; Dumberry and Rivoldini, 2015). Such compositional stratification induced by Fe-snow has also been suggested to occur in the cores of Ganymede and Mars (e.g., Ruckriemen et al, 2015, 2018; Davies and Pommier, 2018). In this study, however, we concentrate exclusively on the behaviour and implications of thermal stratification in the upper part of the cores of Earth and Mercury.

In the simplest scenario that the liquid planetary core is well-mixed and that stratification of an upper liquid core layer is exclusively thermal in nature, the temperature profile satisfies the conduction equation in the stratified region of the core and it approaches the adiabat in the lower region of the core where heat fluxes are super-adiabatic. Thermal evolution modelling of a core that is partly conductive and partly adiabatic involves a moving boundary problem in which the radius of the interface between the conductive and adiabatic regions varies with time. Assuming that thermal boundary layers can safely be neglected in the liquid core (Jaupart et al., 2015), two boundary conditions apply at the interface between the adiabatic and conductive regions. One condition is that the temperature at this interface is continuous. The other condition is that the flux of energy is continuous at the interface, which for a continuous conductivity implies that the temperature gradient at the interface is equal to the adiabatic gradient. Combined with the principle of energy-conservation, these two boundary conditions determine the temperature and the location of the interface.

Gubbins et al. (1982) studied thermal stratification in Earth's liquid core assuming a constant temperature of the core-mantle boundary. Labrosse et al. (1997) used a heat flux boundary condition at the core-mantle boundary that decreases linearly with time. The latter study reformulates the conduction problem with a moving boundary as a Stefan problem and solves it by conventional numerical methods. The growth-rate and consequences of the thermally stratified region in the upper part of the core is largely controlled by the evolution of the heat flux at the core-mantle boundary. In the case-study presented by Labrosse et al. (1997; their figures 4 and 5), thermal stratification initiates at about 1.5 billion years ago, and the stratified region extends to almost 600 km below the core-mantle boundary at present. They show that the influence of such a recent thermally stratified on the thermal evolution of Earth's core is small. For this reason, thermal stratification is commonly neglected in thermal evolution modelling for the Earth for which thermal stratification itself is not the main interest (e.g. Davies et al., 2015; Labrosse, 2015). However, the thermal coupling between the mantle and the thermal stratification in Earth's core has never been considered, because the temperature (Gubbins et al., 1982) or heat flux (Labrosse et al., 1997; Lister and Buffet, 1998) at the core-mantle boundary was always strictly imposed.

Knibbe and van Westrenen (2018) is the only thermal evolution study that considered a thermally stratified region in the outer core of Mercury. They refer to the analytical temperature profile that they adopted in the conductive region as a 'steady-state' solution of the conduction equation, although the solution implicitly considers secular cooling (or secular heating) of the conductive domain and therefore does not imply a steady-state. The adopted temperature profile is more appropriately referred to as the 'steady-flux' solution of the conduction equation, because the adopted analytical temperature profile is the solution of the conduction equation in scenarios with fluxes that are constant in time. The implementation of this steady-flux temperature profile in an energy-conserved thermal evolution approach for the core yields a considerable simplification compared to adopting a numerical solution of the conduction equation as is done by Labrosse et al. (1997), because it does not require discretisation of the conductive region. Knibbe and van Westrenen (2018) derived the heat flux at the core-mantle boundary from the temperature profile in the mantle, of which a parametrized thermal behaviour was coupled to the core. The results of Knibbe and van Westrenen (2018) indicate that thermal stratification in Mercury's core strongly influences the thermal evolution of both Mercury's core and mantle. Because the flux boundary conditions of the thermally stratified region in a planetary core, however, do vary with time, the steady-flux temperature profile is not the exact solution to the time-variable conduction problem that applies in this context.

In this study, we further develop and refine the procedure of Knibbe and van Westrenen (2018) for thermal evolution modelling of a terrestrial core with a thermally stratified upper layer in the liquid core and apply this method to the Earth and Mercury. For sake of simplicity, compositional fluxes are not considered in this study and the thermal conductivity is assumed constant. For the Earth, we study the scenario that is described above in which thermal stratification occurs after the onset of inner core solidification, and we investigate the thermal coupling between the core and the mantle in light of the possible thermal stratification in the upper part of the core. For Mercury, we validate the conclusions of Knibbe and van Westrenen (2018) that were obtained by using the steady-flux assumption.

The energy-conserved approach to modelling the thermal evolution of the core is outlined in section 2. Motivated by the steady-flux solution that is used by Knibbe and van Westrenen (2018), a piece-wise steady-flux (PWSF) numerical scheme is developed that adopts a steady-flux form of the temperature profile in each discretization interval (section 3.1). We show that the PWSF numerical scheme quickly converges to analytical solutions of standard conduction problems on a line segment (Appendix A) and in the radial direction of a sphere (section 3.1). The PWSF numerical scheme is also tested on a spherically symmetric shell of fixed size with time-variable boundary conditions that more closely resembles an outermost stratified layer of Mercury's core. Because an analytical solution does not exist for this more realistic case, the benchmark model for this scenario is provided by a conventional central finite difference scheme with a sufficiently high resolution (section 3.1). The PWSF numerical scheme is implemented in an energy-conserved approach for thermal evolution modelling of a core with a thermally stratified upper region (section 3.2). The case study of Labrosse et al. (1997), which considers thermal stratification at the top of Earth's core by a linear decrease with time of the core-mantle boundary heat flux, is used to validate this approach. Thermal evolution scenarios for the Earth (section 4.1) and for Mercury (section 4.2) are developed and performed, to test the performance of the numerical method and examine the implications of thermal stratification in the core for the thermal evolution of these planets. A summary ends the paper (section 5).

## 2 Thermal evolution modelling with thermal stratification at the top of the core

We assume that stratification initiates at the core-mantle boundary with radial coordinate $R_c$ and consider a spherically symmetric temperature distribution of the planetary core that is separated in an upper conductive and lower adiabatic region. We write the core's temperature profile at time $t$ as

$$T(t,r) = \begin{cases} T_{cd}(t,r); & R_a(t) < r < R_c \\ T_{ad}(t,r); & 0 < r < R_a(t) \end{cases}, \quad (1)$$

where $R_a$ denotes the radial coordinate of the interface between the conductive region and the adiabatic region of the core, and $T_{cd}$ and $T_{ad}$ denote temperature in the conductive and adiabatic regions, respectively. The energy of the core $(E_c)$ at time $t$ is given by

$$E_c(t) = E_{ad}(t) + E_{cd}(t) = \rho c \int_0^{R_a(t)} 4\pi r^2 T_{ad}(t,r)dr + \rho c \int_{R_a(t)}^{R_c} 4\pi r^2 T_{cd}(t,r)dr \ , \quad (2)$$

with $E_{ad}$ and $E_{cd}$ the energy of the adiabatic region and of the conductive region, respectively, $\rho$ the density of the core, and $c$ the heat capacity of the core per unit mass, which are both assumed constant. The time-variation of the core's energy is controlled by the heat loss to the mantle and heat production in the core's interior

$$\frac{dE_c(t)}{dt} = -F_c(t) 4\pi {R_c}^2 + Q(t), \quad (3)$$

with $F_c$ the heat flux per unit surface area at the core-mantle boundary and $Q$ the combined rate of latent and gravitational energy production that results from solidification of an inner core.

In the development of the heat-balance equation at the interface, it is convenient to introduce $\varepsilon(r)$ as the ratio of the volume-averaged temperature in the adiabatic region and the local temperature at a radius within the adiabatic region

$$\varepsilon(t,r) = \frac{\int_0^r 4\pi x^2 T_{ad}(t,x)dx}{\frac{4\pi}{3} r^3 T(t,r)}, r < R_a(t). \quad (4)$$

Because the energy interior to a radius $r$ is related to the temperature profile up to that radius, we can write the temperature at the interface $(T_a)$ as

$$T_a(t) = \frac{E_a(t)}{\varepsilon\big(t, R_a(t)\big) \rho c \frac{4\pi}{3} R_a(t)^3}, \quad (5)$$

If the interface radius is constant in time, the energy balance equation restricted to the adiabatic region is

$$\frac{dE_a(t)}{dt} = -F_a(t) 4\pi R_a(t)^2 + Q(t), \quad (6)$$

where the heat flux out of the adiabatic region is given by

$$F_a(t) = -k \frac{\partial T_{ad}\big(t, R_a(t)\big)}{\partial r}, \quad (7)$$

with $k$ the thermal conductivity of the core. In the scenarios we will study in this paper, the adiabatic gradient is assumed to be proportional to temperature. In section 3.2, we will show that the time-dependence of $\varepsilon(t,r)$ vanishes under this assumption, and we can write $\varepsilon(r) := \varepsilon(t,r)$. The time derivative of the interface temperature then only depends on time trough $Q(t)$, $F_a(t)$ and the rate of change of $R_a$. We take the radial variation of $R_a$ into consideration by integrating the adiabat gradient, and can write the time derivative of the interface temperature as (Knibbe and van Westrenen, 2018)

$$\frac{dT_a(t)}{dt} = \frac{Q(t) - F_a(t) 4\pi R_a(t)^2}{\varepsilon(R_a(t))\rho c \frac{4\pi}{3} R_a(t)^3} + \frac{dR_a(t)}{dt} \frac{\partial T_{ad}\big(t, R_a(t)\big)}{\partial r}. \quad (8)$$

Equations (3), (6) and (8) implicitly incorporate a certain amount of secular cooling of the inner core that results from varying the adiabatic temperature profile with time from energy-balance considerations. These equations do not consider the possibility that the conductive cooling rate of the inner core can be larger than that. This conductive cooling of the inner core is, for example, considered in Labrosse et al. (1997), but is neglected in this paper because it does not play an important role in the evolution of Earth's core and in case studies of this paper (as we will show in section 3.2). Additionally, equations (6) and (8) assume that the internal heat production by solidification of the inner core takes place in the adiabatic region of the core, which is true in the scenarios that we will consider in this study where the thermally stratified region does not extend to the solid inner core.

In order to determine the core's temperature profile and the interface radius $R_a$ at time $t + dt$ from a temperature profile and interface radius that are known at time $t$, we proceed as follows: We first construct the adiabatic temperature profile and the conductive temperature profile at time $t + dt$ as a function of the, a priori unknown, $R_a(t + dt)$. Next, we solve for $R_a(t + dt)$ by forcing continuity of temperature at the interface $\big(T_{ad}(t + dt, R_a(t + dt)) = T_{cd}(t + dt, R_a(t + dt))\big)$. We perform the first step by forward integrating the energy of the core with respect to time (equation 3) to obtain $E_c(t + dt)$, and by forward integrating equation (5) with respect to time for a chosen $dR_a(t)/dt$ (or $R_a(t + dt)$) to obtain $T_a(t + dt)$. The energy of the adiabatic region can be determined from anchoring the adiabat at the interface: $T_{ad}\big(t + dt, R_a(t + dt)\big) = T_a(t + dt)$. Equation (2), evaluated at time $t + dt$, then provides an energy-conservation constraint for the conductive temperature profile at time $t + dt$, in terms of $R_a(t + dt)$. This energy constraint together with the continuity of flux, which is at the core-mantle boundary given by

$$\frac{\partial T_{cd}(t + dt, R_c)}{\partial r} = -\frac{F_c}{k} \qquad (9)$$

and at the interface $R_a(t + dt)$ given by

$$\frac{\partial T_{cd}\big(t + dt, R_a(t + dt)\big)}{\partial r} = \frac{\partial T_{ad}\big(t + dt, R_a(t + dt)\big)}{\partial r}, \qquad (10)$$

constrain the conductive temperature profile in the domain $R_a(t + dt) < r < R_c$ fully in terms of $R_a(t + dt)$. With a formulation of the conductive temperature profile for these energy and boundary constraints, as explained below, we solve $R_a(t + dt)$ by imposing continuity of temperature at the interface

$$T_{cd}\big(t, R_a(t + dt)\big) = T_{ad}\big(t, R_a(t + dt)\big). \qquad (11)$$

The conductive temperature profile satisfies the equation for radial conduction through a sphere, which reads

$$\rho c \frac{\partial T_{cd}(t,r)}{\partial t} = \frac{1}{r^2}\frac{\partial}{\partial r}\left(r^2 k \frac{\partial T_{cd}(t,r)}{\partial r}\right) \qquad (12)$$

for $R_a < r < R_c$. The right-hand side of equation (12) is the Laplacian of the spherically symmetric temperature profile. The procedure outlined above requires a solution for the conductive temperature profile for which the temperature and the radial derivative (equation 7) can be computed at $R_a(t + dt)$, which is a priori not known. It is therefore convenient that the solution of the conduction equation (12), which is approximated here, is continuous and differentiable with respect to radius. Knibbe and van Westrenen (2018) adopted the steady-flux solution of the conduction equation for the conductive region

$$T_{cd}(t,r) = -\frac{S(t)}{6k} r^2 + \frac{A(t)}{r} + B(t), \qquad (13)$$

which is continuous and differentiable with respect to radius. The time-dependent functions in equation (13) are defined by

$$S(t) = k \frac{4\pi \left(\frac{\partial T(t, R_a(t))}{\partial r} R_a(t)^2 - \frac{\partial T(t, R_c)}{\partial r} {R_c}^2\right)}{\frac{4}{3}\pi\big({R_c}^3 - R_a(t)^3\big)}, \qquad (14)$$

$$A(t) = -\frac{1}{k}\left(k \frac{\partial T(t, R_c)}{\partial r} {R_c}^2 + \frac{S(t)}{3} {R_c}^3\right), \qquad (15)$$

and

$$B(t) = \frac{E_{cd}(t) + \rho c\left(\frac{4\pi}{30k}S(t)\left(R_c{}^5 - R_a(t)^5\right) - 2\pi A(t)\left(R_c{}^2 - R_a(t)^2\right)\right)}{\rho c\frac{4}{3}\pi\left(R_c{}^3 - R_a(t)^3\right)}. \quad (16)$$

The temperature profile of equation (13) satisfies the conduction equation locally only if the thermal gradient boundary conditions are constant in time (therefore refer to it as the steady-flux solution), and is realized only in the limit of time to infinity (or if the initial temperature profile is of steady-flux form). A particular feature of this solution is that the cooling rate $\partial T_{cd}(t,r)/\partial t$ is uniform in space if the boundary conditions are constant in time (then, $S(t)$ and $A(t)$ are constant in time and $B(t)$ is linear in time). The interested reader can see in Appendix A, where conduction on a line-segment is considered, that the general solution (equation A17) indeed tends to the steady-flux solution. Because the flux boundary conditions vary in time in the cores of Earth and Mercury, the steady-flux solution is not the exact solution to the problem at hand. The steady-flux solution, nevertheless, satisfies equation (12) on-average across the conductive domain, in the sense that

$$\frac{1}{\frac{4\pi}{3}\left(R_c{}^3 - R_a{}^3\right)}\int_{R_a}^{R_c} 4\pi r^2 \rho c \frac{\partial T_{cd}(t,r)}{\partial t}\,dr = \frac{1}{r^2}\frac{\partial}{\partial r}\left(r^2 k \frac{\partial T_{cd}(t,r)}{\partial r}\right). \quad (17)$$

By computing either side of equation (17) by using equation (13), we infer that this volumetrically averaged rate of secular heating of the conductive region (left hand-side) and the Laplacian of the temperature profile (right-hand side) are equal to $-S(t)$. Hence, $S$ is the average secular cooling, or the average heat loss per units of volume and time, of the conductive region.

Instead of applying the temperature profile given by equation (13) as in Knibbe and van Westrenen (2018), we here develop a piece-wise steady-flux (PWSF) numerical scheme that assumes the steady-flux temperature profile in intervals of the discretised conductive domain and that is continuous and differentiable with respect to radius. The numerical scheme ensures that equation (12) is satisfied on-average in each of the intervals, in the sense of equation (17) with the integral performed over the interval. Therefore, the parameters $S_i$ are locally equal to the negative of the Laplacian of the temperature profile within interval $i$, and are also equal the average secular cooling of that interval. By increasing the number of intervals, the PWSF scheme approximates the conduction equation at increasingly local scale.

## 3 The piecewise steady flux (PWSF) scheme

### 3.1 The PWSF numerical scheme for radial conduction in a sphere or spherical shell

We recall that radial conduction in a sphere or spherical shell is described by

$$\frac{\partial T(t,r)}{\partial t} = \frac{1}{r^2}\frac{\partial}{\partial r}\left(r^2 \kappa \frac{\partial T(t,r)}{\partial r}\right), \qquad r_0 < r < r_N, t > 0, \qquad (18)$$

with $\kappa = k/(\rho c)$ the thermal diffusivity. We develop the PWSF numerical scheme that approximates the solution of equation (18). This development is analogous to the PWSF scheme for conduction on a line segment that is described and tested in Appendix A. The domain $(r_0\ r_N)$ is discretized in $N$ intervals $(r_{i-1}\ r_i)$ for $i = 1, 2, \dots N$ with length $L_i = r_i - r_{i-1}$ and volume $V_i = (4\pi/3) \cdot ({r_i}^3 - {r_{i-1}}^3)$. It is convenient to introduce

$$U(t) = \begin{bmatrix} U_1(t) \\ U_2(t) \\ \vdots \\ U_N(t) \end{bmatrix} = \begin{bmatrix} \int_{r_0}^{r_1} T(t,r) 4\pi r^2 dr \\ \int_{r_1}^{r_2} T(t,r) 4\pi r^2 dr \\ \vdots \\ \int_{r_{N-1}}^{r_N} T(t,r) 4\pi r^2 dr \end{bmatrix}, \qquad (19)$$

the array of dimension $N$ of the volumetric integrals of $T(t,r)$ over the intervals. By integration of equation (18) over $r$, we obtain

$$\frac{dU(t)}{dt} = 4\pi\kappa \left( \begin{bmatrix} \frac{\partial T(t,r_1)}{\partial r}{r_1}^2 \\ \frac{\partial T(t,r_2)}{\partial r}{r_2}^2 \\ \vdots \\ \frac{\partial T(t,r_N)}{\partial r}{r_N}^2 \end{bmatrix} - \begin{bmatrix} \frac{\partial T(t,r_0)}{\partial r}{r_o}^2 \\ \frac{\partial T(t,r_1)}{\partial r}{r_1}^2 \\ \vdots \\ \frac{\partial T(t,r_{N-1})}{\partial r}{r_{N-1}}^2 \end{bmatrix} \right), \qquad (20)$$

which we integrate with respect to time by a forward finite volume scheme. In equation (20), the thermal gradients $\partial T(t,r_0)/\partial r$ and $\partial T(t,r_N)/\partial r$ are given by boundary conditions and the thermal gradients $\partial T(t,r_i)/\partial r$ with $i = 1, 2, \dots\ N-1$ need to be estimated. The $N-1$ unknown temperature gradients can be approximated by a finite difference of the ratio of the temperature difference between the midpoints of each interval over the distance between such midpoints

$$\frac{\partial T(t,r_i)}{\partial r} = \frac{2\left(T\left(t, r_i + \frac{1}{2}L_{i+1}\right) - T\left(t, r_i - \frac{1}{2}L_i\right)\right)}{L_{i+1} + L_i}. \qquad (21)$$

If intervals are of equal length ($L_i = dr$) and temperature is implicitly assumed constant across any interval, equations (19-21) are equivalent to a finite volume numerical scheme for the conduction equation (equation 18)

$$\frac{T(t+dt,r)-T(t,r)}{dt} = \kappa\frac{4\pi}{V_i}\frac{{r_{i-1}}^2\left(T\left(t,r_{i-1}-\frac{1}{2}dr\right)-T\left(t,r_{i-1}+\frac{1}{2}dr\right)\right)-{r_i}^2\left(T\left(t,r_i-\frac{1}{2}dr\right)-T\left(t,r_i+\frac{1}{2}dr\right)\right)}{dr^2}\ ;\quad r_{i-1} < r < r_i\ \ ,(22)$$

which is based on forward first-order finite differences with respect to the time and central differences with respect to radius, and is also known as the method of lines. Such a scheme, however, is not accompanied by a continuous and differentiable formulation of temperature, which would be convenient for the thermal evolution approach that is outlined in section 2. Instead of using finite differences (equation 21) to estimate the thermal gradients at the interval boundaries, the PWSF numerical scheme assumes a steady-flux form of the temperature profile $T_i(t,r)$ across interval $i$

$$T_i(t,r) = -\frac{S_i(t)}{6\kappa}r^2 + \frac{A_i(t)}{r} + B_i(t),\ \ r_{i-1} < r < r_i, i = 1,2,\dots,N, \qquad (23)$$

which is piece-wise continuous and differentiable, with

$$S_i(t) = \kappa\frac{4\pi\left(\frac{\partial T(t,r_{i-1})}{\partial r}{r_{i-1}}^2 - \frac{\partial T(t,r_i)}{\partial r}{r_i}^2\right)}{V_i}, \qquad (24)$$

$$A_i(t) = -\frac{1}{\kappa}\left(\kappa\frac{\partial T(t,r_i)}{\partial r}{r_i}^2 + \frac{S_i(t)}{3}{r_i}^3\right), \qquad (25)$$

and

$$B_i(t) = \frac{U_i + \frac{4\pi}{30\kappa}S_i(t)({r_i}^5 - {r_{i-1}}^5) - 2\pi A_i(t)({r_i}^2 - {r_{i-1}}^2)}{V_i}. \qquad (26)$$

The $N-1$ thermal gradients are at the interval boundaries are determined by imposing continuity at the interval boundaries ($T_i(t,r_i) = T_{i+1}(t,r_i)$ for $i = 1,2,\dots\ N-1$). This gives rise to $N-1$ equations that can be written as

$$My = b, \qquad (27)$$

with

$$y = \begin{bmatrix} \frac{\partial T(t,r_1)}{\partial r} \\ \frac{\partial T(t,r_2)}{\partial r} \\ \vdots \\ \frac{\partial T(t,r_{N-1})}{\partial r} \end{bmatrix}, \qquad (28)$$

$b$

$$= \begin{bmatrix} \frac{U_2(t)}{V_2} - \frac{U_1(t)}{V_1} + \frac{\partial T(t,r_0)}{\partial r}\left(\frac{2\pi r_0^2 r_1^2}{V_1} - \frac{\frac{16}{30}\pi^2 r_0^2(r_1^5 - r_0^5) + \frac{8}{3}\pi^2 r_0^2 r_1^3(r_1^2 - r_0^2)}{V_1^2}\right) \\ \frac{U_3(t)}{V_3} - \frac{U_2(t)}{V_2} \\ \vdots \\ \frac{U_{N-1}(t)}{V_{N-1}} - \frac{U_{N-2}(t)}{V_{N-2}} \\ \frac{U_N(t)}{V_N} - \frac{U_{N-1}(t)}{V_{N-1}} + \frac{\partial T(t,r_N)}{\partial r}\left(\frac{2\pi r_N^2 r_{N-1}^2 + 6\pi r_N^2(r_N^2 - r_{N-1}^2) + 4\pi \frac{r_N^5}{r_{N-1}}}{3V_N} - \frac{r_N^2}{r_{N-1}} - \frac{\frac{16}{30}\pi^2 r_{N-1}^2(r_N^5 - r_{N-1}^5) + \frac{8}{3}\pi^2 r_N^5(r_N^2 - r_{N-1}^2)}{V_N^2}\right) \end{bmatrix}, \quad (29)$$

and $M$ a tridiagonal matrix with

$$M_{i,i-1} = -\frac{2\pi r_{i-1}^2 r_i^2}{V_i} + \frac{\frac{16}{30}\pi^2 r_{i-1}^2(r_i^5 - r_{i-1}^5) + \frac{8}{3}\pi^2 r_{i-1}^2 r_i^3(r_i^2 - r_{i-1}^2)}{V_i^2};\ i = 2,3,\ldots,N-1, \quad (30)$$

$$M_{i,i} = -r_i + \frac{2\pi r_i^4 + 2\pi r_i^2(r_i^2 - r_{i-1}^2)}{V_i} - \frac{\frac{16}{30}\pi^2 r_i^2(r_i^5 - r_{i-1}^5) + \frac{8}{3}\pi^2 r_i^5(r_i^2 - r_{i-1}^2)}{V_i^2} - \frac{\frac{16}{30}\pi^2 r_i^2(r_{i+1}^5 - r_i^5) + \frac{8}{3}\pi^2 r_i^2 r_{i+1}^3(r_{i+1}^2 - r_i^2)}{V_{i+1}^2} + \frac{2\pi r_i^4 + 4\pi r_i r_{i+1}^3}{3V_{i+1}};\ i = 1,2,\ldots,N-1, \quad (31)$$

and

$$M_{i,i+1} = -\frac{2\pi r_{i+1}^2 r_i^2 + 6\pi r_{i+1}^2(r_{i+1}^2 - r_i^2) + 4\pi \frac{r_{i+1}^5}{r_i}}{3V_{i+1}} + \frac{r_{i+1}^2}{r_i} + \frac{\frac{16}{30}\pi^2 r_{i+1}^2(r_{i+1}^5 - r_i^5) + \frac{8}{3}\pi^2 r_{i+1}^5(r_{i+1}^2 - r_i^2)}{V_{i+1}^2};\ i = 1,2,\ldots,N-2. \quad (32)$$

The $N-1$ thermal gradients $\partial T(t, r_i)/\partial r$ are solved using equation (27) for given $U$ at time $t$, which implicitly assures differentiability of temperature at the interval boundaries. The obtained thermal gradients are used to integrate $U$ forward in time (equation 20). The temperature profile can be reconstructed by equations (23-26). In the remainder of this section, we show by testing that this PWSF scheme solves the conduction equation up to arbitrary precision with increase of numerical resolution.

[Place of figure 1]

We perform forward integrations of the PWSF numerical scheme with $N = 16$, $N = 4$ and $N = 1$ on a sphere ($r_0 = 0$) with initial condition $T(0, r) = 0$ and with inhomogeneous Neumann boundary conditions that are constant in time

$$\frac{\partial T(t, r_0)}{\partial r} = 0, \frac{\partial T(t, r_N)}{\partial r} = g_N. \qquad (33)$$

The analytical solution of this problem is given by (Crank, 1956)

$$T(t, x) = r_N g_N \left( \frac{3\kappa t}{{r_N}^2} + \frac{r^2}{2{r_N}^2} - \frac{3}{10} - \frac{2r_N}{r} \sum_{n=1}^{\infty} \frac{\sin(\lambda_n r)}{{\lambda_n}^2 {r_N}^2 \sin(\lambda_n r_N)} e^{-\kappa t {a_n}^2} \right), \quad (34)$$

where the $\lambda_n$ are the positive roots of

$$\lambda_n r_N \cdot \cot(\lambda_n r_N) = 1. \qquad (35)$$

Because a thermal gradient is imposed at the surface of the sphere, time-variation of temperature (the rate of heating or cooling) propagates from the surface of the sphere to the inner part of the sphere. We recall that the conduction equation (12) relates the heating (or cooling) rate to the Laplacian of the temperature profile, which for the PWSF numerical scheme is piecewise constant over each interval (equal to $-S$). Additionally, the PWSF scheme ensures conservation of energy on individual intervals by equation (20). For these reasons, the PWSF numerical scheme approximates the solution of the conduction equation with high accuracy if the adopted intervals are smaller than the spatial scale of variations in the heating (or cooling) rate. For example, at $t = \tau/4^5$, with $\tau = {r_N}^2/\kappa$ the conduction timescale, the heating (or cooling) rate is significant only in the small outermost region of the sphere with $r/r_N > 0.9$ (figure 1a). Differences between the analytical solution of the conduction equation and the temperature profile of the PWSF schemes with $N = 4$ and $N = 1$ are, therefore, significantly in error whereas the errors of the PWSF scheme with $N = 16$ are small, at $t = \tau/4^5$ (figure 1a). As time progresses with the constant boundary condition, the heating (or cooling) rate distributes across the conductive domain and the piece-wise constant Laplacian of temperature of the low resolution PWSF schemes more appropriately approximates to the local cooling rate. The differences between the analytical solution and the temperature profile of the PWSF scheme $N = 4$ are small since $t \sim \tau/4^3$ (figure 1b). Since $t \sim \tau/4$, radial variations in the cooling or heating rate across the conductive domain have practically disappeared and also the PWSF scheme with $N = 1$ describes the temperature profile and the thermal evolution relatively accurately (figure 1c).

We remark that the precision of the numerical scheme can be improved by taking smaller intervals in the region where large variations in the heating (or cooling) rate exist. For example, if the thermal gradient boundary condition at the outer radius varies with time, one may discretize the sphere in intervals of equal volume rather than in intervals of equal thickness to reduce the interval size in the outer part of the sphere.

[Place of figure 2]

Having evaluated the characteristics of the PWSF numerical scheme on a sphere with fixed Neumann boundary conditions, we now assess its performance on a spherical shell with time-variable boundary fluxes. We consider a conductive layer of 500 km thick in Mercury's 2000 km radius core, which is a possible size for the thermally stratified layer in Mercury's core (Knibbe and van Westrenen, 2018). Assuming a core density of 7200 kg·m$^{-2}$, heat capacity of 840 J·kg$^{-1}$·K$^{-1}$ and thermal conductivity of 50 W·m$^{-1}$·K$^{-1}$, the conduction timescale ($\tau = (r_N - r_0)^2/\kappa$) is about 1 billion years. We showed that the conduction solution, assuming fixed boundary conditions, converges to the steady-flux form on a timescale of $\tau/4$ (figure 1c and figure A1 of Appendix A), which is equivalent to ~250 million years for a 500 km thick spherical shell in Mercury's core. Various thermal evolution studies of Mercury show that the core-mantle boundary heat flux decreases rapidly in the earliest several hundred million years, whereas the heat flux at the core-mantle boundary remains relatively stable in the planet's later evolution (e.g., Tosi et al., 2013; Knibbe and van Westrenen, 2018). We therefore test the PWSF numerical scheme with $N = 1$, $N = 2$, and $N = 4$ using flux boundary conditions that vary strongly within ~200 million years. The initial condition is set by a steady-flux temperature profile (equation 12, grey line in figure 2), derived from an initial thermal gradient $\partial T(t, r_N)/\partial r$ of -6.5·10$^{-4}$ K·m$^{-1}$ at the outer radius $r_N = 2000$ km, a thermal gradient of -1.5·10$^{-4}$ K·m$^{-1}$ at the inner radius $r_0$ =1500 km, and a volume-average temperature of 1900 K of the spherical shell. We vary the thermal gradient at $r_N$ linearly with time to -1.5·10$^{-4}$ K·m$^{-1}$ between $0 < t < \tau/5$ and keep it constant at this value for $t > \tau/5$. Because no analytical solution exists for this problem, a benchmark evolution is generated by using the finite difference approximation for the thermal gradients (equation 21) with $N = 100$. The obtained benchmark solution does not significantly improve by further increase of numerical resolution. For the PWSF schemes, we discretised the spherical shell in intervals of equal volume.

The radial variations in the cooling rate increase in the period $0 < t < \tau/5$ (figures 2a and 2b), during which the thermal gradient at Mercury's core-mantle boundary decreases with time (in absolute sense). At $t = \tau/5$, the PWSF scheme with $N = 1$ has ~24 K higher temperature at $r_N$ and ~26 K lower temperature at $r_0$ as compared to the benchmark evolution. The PWSF scheme with $N = 2$ and with $N = 4$ have temperature errors of ~4 K, and ~1 K, respectively, at the boundaries of the spherical shell. The radial variations in the cooling rate rapidly decrease in the period $\tau/5 < t < 2\tau/5$ in which boundary conditions are constant over time (figure 2c). At $t = 2\tau/5$, the temperature differences between the PWSF schemes with $N = 1$, $N = 2$, or $N = 4$ and the benchmark model are reduced to below 4 K, below 1.5 K or below 0.5 K, respectively. As time continuous, all PWSF schemes converge to the benchmark model.

In the period where the heat flux at the core-mantle boundary strongly decreases with time, the outermost region of the spherical shell loses less energy than the heat that is conducted to it from lower regions of the shell. This results in a temporary and local negative cooling rate, which thus represents secular heating, of a region just below the core-mantle boundary (figure 2b).

### 3.2 Implementation of the PWSF scheme in a thermal evolution model of a planetary core

We implement the PWSF numerical scheme (section 3.1) in the approach for thermal evolution modelling of a planetary core with time-variable conductive domain, which is outlined in section 2. We discretize the core in $N$ intervals of equal volume, and number the intervals starting from 1 for the uppermost interval below the core-mantle boundary to $N$ at the planet's center. Accordingly, the radius of the lower boundary of interval $i$ is given by

$$\tilde{R}_i = \left(\frac{(N-i){R_c}^3}{N}\right)^{1/3}, i = 1, 2, \dots (N-1). \quad (36)$$

For as long as $R_a > \tilde{R}_1$, the PWSF formulation of the conductive temperature profile is equivalent to the steady-flux temperature profile given by equation (13). For the case that $\tilde{R}_j > R_a > \tilde{R}_{j+1}$ with $j > 0$, we extend the scheme that is presented in section 2 as follows. As before, we integrate the energy equations (3) and (8) forward in time to determine $Ec(t+dt)$ and $T_a(t+dt)$ as a function of $R_a(t+dt)$. In the total energy of the core, we separate the energy of the interval that is partially conductive:

$$\rho c \int_{R_a(t)}^{\tilde{R}_j} 4\pi r^2 T_{cd}(t,r)dr = E_c(t) - \sum_{i=1}^{j} E_{cd,i}(t) - E_{ad}(t), \quad (37)$$

where the $E_{cd,i}$ for $i$ from 1 to $j$ denote the energies of the $j$ intervals that are fully conductive and $E_{ad}$ denotes the energy of the adiabatic region. Energy $E_{cd,i}$ of the $i$'th fully conductive interval is integrated forward in time by (see equation (20) for $U$)

$$\frac{dE_{cd,i}}{dt} = 4\pi k \left(\frac{\partial T(t,\tilde{R}_{i-1})}{\partial r}{\tilde{R}_{i-1}}^2 - \frac{\partial T(t,\tilde{R}_i)}{\partial r}{\tilde{R}_i}^2\right). \quad (38)$$

Because the energy of the adiabatic region is also known as a function of $R_a(t+dt)$ (see below), the energy of the conductive part of the partially conductive interval can then be determined at time $t+dt$ from equation (37), evaluated at time $t+dt$. These energies $(E_{cd,i})$ are used to determine the thermal gradients $\partial T(t+dt,\tilde{R}_i)/\partial r$ for $i$ from 1 to $j$ by applying equations (27-32) on the domain $R_a(t+dt) < r < R_c$. Boundary flux conditions for equations (27-32) are set by the core-mantle boundary heat flux and by the adiabatic gradient (equations 7 and 9). The temperature profile ($T_{cd}(t+dt,r)$) of the conductive shell is then given by equations (23-26). Similar to the procedure outlined in section 2.1, the radius $R_a(t+dt)$ is found by imposing continuity of temperature at the interface

$$T_{cd}\big(R_a(t+dt)\big) = T_a(t+dt) = T_{ad}\big(R_a(t+dt)\big). \quad (39)$$

At each first timestep that $R_a$ becomes smaller than $\tilde{R}_j$ for some $j > 0$, a new interval becomes fully conductive ($j$ increases by one). At each first timestep that $R_a > \tilde{R}_j$ for some $j > 0$, the lowermost fully conductive interval becomes partially conductive ($j$ decreases by one).

We parametrize the adiabatic temperature profile in the core as

$$T_{ad}(t,r) = T_0(t)\left(1 + \sum_{i=1}^{n}\left(T_i r^i\right)\right), \qquad (40)$$

with $T_i$ for $i$ from 0 to $n$ parameters that describe the adiabat. The relative volume-averaged temperature in an adiabatic sphere of radius $r$ with respect to the local temperature at $r$ is then given by

$$\varepsilon(t,r) = \frac{\int_0^r 4\pi x^2 T_{ad}(t,x)dx}{\frac{4\pi}{3} r^3 T(t,r)} = \frac{1 + \sum_{i=1}^{n} \frac{3}{3+i} T_i r^i}{1 + \sum_{i=1}^{n} T_i r^i} \quad ; r \leq R_a(t). \qquad (41)$$

Note that, because the adiabatic temperature at radius smaller than $r$ is proportional to temperature at radius $r$ (equation 40), the dependence of $\varepsilon(t,r)$ to $T(t,r)$ vanishes (see equation 41). Because the time-dependence of $\varepsilon(t,r)$ is through the time-dependence of temperature, the time-dependence of $\varepsilon$ disappears ($\varepsilon(r) \coloneqq \varepsilon(t,r)$, for any $t$ and any $r \leq R_a$). Therefore, the energy of the adiabatic part of the core (used in equation 8) can be written as

$$E_{ad}(t) = T_a(t)\varepsilon\big(R_a(t)\big)\rho c \frac{4\pi}{3} R_a(t)^3. \qquad (42)$$

[Place of figure 3]

We validated this thermal evolution approach by applying it to the thermal evolution scenario of Labrosse et al. (1997) that considers thermal stratification in an upper region of Earth's core. The set-up of this thermal evolution scenario is described in Appendix B. As is outlined in section 3.2, the core is discretised in $N$ intervals of equal volume, which effectively increases the numerical resolution of the PWSF scheme for the thermally stratified layer only if the uppermost intervals are smaller than the thermally stratified layer. To evaluate the improvement of the numerical scheme with the increase of numerical resolution, thermal evolution runs are therefore performed with $N = 1$, with $N = 7$, and with $N = 14$. For comparison, we additionally performed a thermal evolution model that described the core's temperature profile by the adiabat throughout Earth's evolution, which thus neglects thermal stratification.

The results (figure 3) are very similar to the results presented by Labrosse et al. (1997; their figures 4 and 5). The obtained present-day thickness of the thermally stratified region is 534 km, 566 km, and 569 km for the thermal evolution run with $N = 1$, with $N = 7$, and with $N = 14$, respectively. Labrosse et al. (1997) obtained an almost 600 km thick conductive region at present-day, and do not mention the exact thickness that is obtained. A small difference between the results of Labrosse et al. (1997) and our method is expected because we fitted our parametrization of the adiabat to the adiabat used by Labrosse et al. (1997) (appendix B). Additionally, we describe the temperature profile of the inner core by the adiabat, such that the cooling rate is limited to the rate at which adiabatic temperature profile decreases with time. Labrosse et al. (1997) solved the conduction equation in the inner core, such that the thermal gradient in the inner core decreases with time and the solid inner core cools at a slightly larger rate. For this reason, our model leaves slightly more energy stored in the inner core and we obtain a slightly smaller secular cooling of the inner core (0.17 TW at present) compared to Labrosse et al. (1997) (~0.25 TW at present). This difference of below 0.1 TW is minor as compared to the heat balance of the convective region of the core (figure 3c), which is dominated by latent heat (~1.54 TW), gravitational energy (~1.17 TW), and secular cooling of the liquid adiabatic outer core (~1.12 TW at present-day, starting from ~6.5 TW at the onset of thermal stratification).

We have seen in prior testing that the accuracy of the PWSF scheme improves with increase of numerical resolution (figures 1 and 2). Therefore, the very minor difference between the thermal evolution runs with $N = 7$ or $N = 14$, and supported by the close similarities with the results of Labrosse et al. (1997), indicates that these numerical schemes closely approximate the actual solution of the thermal evolution scenario. Additional evidence for this conjecture comes from comparing the radial variation of secular cooling across the conductive region of this scenario with that of the test example presented in figure 2. In this scenario, the cooling rate varies maximally by ~24 nW·m$^{-3}$, from ~-12 nW·m$^{-3}$ at the core-mantle boundary to ~12 nW·m$^{-3}$ at the interface radius (figure 3d). In figure 2, we have seen that much larger variations in the cooling rate of ~117 nW·m$^{-3}$ are modelled to about 1 K accuracy by discretizing the conductive spherical shell of 500 km thick into four intervals. The thermal evolution runs of this scenario with $N = 7$ and with $N = 14$ effectively discretize the conductive layer into three and six intervals, respectively (figure 3d), such that the expected deviation from the actual solution of this thermal evolution scenario is indeed very minor.

The PWSF numerical scheme with $N = 1$, which is equivalent to the steady-flux thermal evolution approach of Knibbe and van Westrenen (2018), has the advantage that it is much simpler than those with $N > 1$, because it does not require discretization of the conductive domain. But we observe that the steady-flux assumption for the temperature profile of the thermally stratified region does generate small errors (figure 3). The present-day temperature at the core-mantle boundary of the PWSF scheme with $N = 1$ is 6 K higher than that of the PWSF scheme with $N = 14$. Such a deviation of core-mantle boundary temperature from the actual solution is slightly larger than the ~4 K warmer temperature at the core-mantle boundary presented in figure 2c at $t = 2\tau/5$, which is consistent with the slightly larger radial variations in the cooling rate figure 3d compared to in figure 2c. In figure 2, the temperature errors at the lower boundary of the conductive domain were of similar magnitude as at the core-mantle boundary. However, by implementation of the PWSF scheme with the moving boundary problem, as is done in this scenario (figure 3), the temperature at the interface radius is controlled by the total energy in the adiabatic region (equation 7). For the PWSF scheme with $N = 1$, the deviation from the actual solution near the interface radius is dominantly expressed by the larger size of the adiabatic region (at present 35 km larger in radius than in the PWSF scheme with $N = 14$). Because the adiabatic gradient increases with radius, the PWSF scheme with $N = 1$ effectively integrated a slightly larger heat flux from the adiabatic region into the conductive region than the PWSF schemes with $N = 7$ and $N = 14$, and therefore has about 1.5 K lower temperature in the adiabatic region. The temperature error near the interface (~1.5 K) is thus much smaller than the temperature error at the core-mantle boundary (~6 K). But this temperature error near the interface goes paired with an error in the thickness of the thermally stratified layer (an error of ~35 km in this scenario).

In reproducing this scenario of Labrosse et al. (1997), the temperature at the core-mantle boundary increases in the latest billion years of the evolution in the thermal evolution runs that consider thermal stratification, whereas the adiabatic region cools for the entire duration of the computed thermal evolution (figure 3b). In the thermal evolution run that neglects thermal stratification, the distribution of energy is fixed by the adiabat. Therefore, this thermal evolution run does not take into account that the cooling rate can vary with radius. As a consequence, the thermal evolution run that neglects thermal stratification yields a 67 K lower present-day temperature at the core-mantle boundary, a 4 K higher temperature of the adiabatic region, and a 35 km smaller radius of the inner core than that of the PWSF thermal evolution run with $N = 14$ (figure 3b). The difference in the size of the inner core was also noted by Labrosse et al. (1997) and judged to be of small importance. The difference in core-mantle boundary temperature would, however, influence the heat flux into the mantle. Because this heat flux is the boundary condition for the core's thermal evolution, this interaction with the mantle is expected to provide a feedback for the evolution of the core. This feedback is not considered in this scenario, because the heat flux at the core-mantle boundary is strictly imposed. We will evaluate the effect of that feedback in the next section.

## 4 Thermal stratification in the cores of the Earth and Mercury

## 4.1 Earth

[Place of figure 4]

To investigate the coupling between thermal stratification in the Earth's core and the heat flux at the core-mantle boundary, we develop a thermal evolution scenario in which the heat flux at the core-mantle boundary is determined from the thermal gradient in the lower thermal boundary layer of the mantle. The parametrization of the thermal boundary layer is taken from commonly used parametrized thermal evolution models of mantle convection (e.g. Stevenson et al., 1983; Thiriet et al., 2019) and is described in detail in appendix B. We are interested in the behaviour and implications of thermal stratification in the core, and not specifically in the evolution of the mantle. For this reason, we set the temperature at the top of the lower thermal boundary layer in the mantle ($T_m$) at a constant value of 2630 K, which is the present-day temperature of Earth's lower mantle estimated by Katsura et al. (2010). The parametrization of Earth's core is unchanged relative to the scenario of Labrosse et al. (1997) that is presented in figure 3, to facilitate a direct comparison to that scenario where the heat flux at the core mantle boundary is imposed to decrease linearly with time. Similar to the previous computations, we performed thermal evolution runs with $N = 1$, $N = 7$ and $N = 14$. But the results with $N = 14$ are not shown because differences between the results with $N = 7$ are negligible.

Results of the thermal evolution runs are presented in figure 4. The heat flux at the core mantle boundary initially decreases rapidly with time, but the rate of decrease of the core-mantle boundary heat flux reduces with time (figure 4d). There are three reasons that the withdrawal of energy from the core decreases with time. Firstly, while the temperature at the core-mantle boundary decreases with time, the temperature difference between the core and the mantle decreases and the viscosity of the mantle increases, which lead to a smaller heat flux at the core mantle boundary (figure 4b). Secondly, the rate of decrease of the core-mantle boundary heat flux drops significantly at the onset of core solidification at about 2.7 billion years (figure 4d), due to the production of latent and gravitational energy in the core (figure 4c). And thirdly, after the onset of thermal stratification at the core mantle boundary at about 2.8 billion years, the core mantle boundary cools at a lower rate than the adiabatic region of the core (figure 4b). The reduced cooling rate of the core-mantle boundary, while the thermally stratified layer is developing, leads to a smaller decrease of the core-mantle boundary heat flux.

Because the core-mantle boundary heat flux decreases at a smaller rate in this scenario compared to the scenario of Labrosse et al. (1997) (figure 3), the thermally stratified layer grows at a smaller rate and also the consequences of thermal stratification are reduced. The present-day size of the thermally stratified layer is only 290 km (figure 4a), the core-mantle boundary is only ~12 K warmer at present if thermal stratification is considered compared to if thermal stratification is neglected (figure 4b), and the radius of the inner core is only 10 km larger if thermal stratification is considered compared to if it is neglected (figure 4a). As a result of the small influence of thermal stratification on the temperature of the core-mantle boundary, the present-day heat flux from the core into the mantle is only ~0.18 TW (~3.3%) larger if thermal stratification is accounted for compared to if it is neglected (figure 4c). The latter implies that thermal stratification in the Earth's core does not substantially influence the heat budget of the mantle. We conclude that the consideration of thermal coupling between the core and mantle leads to a smaller growth-rate of the thermally stratified region and also further reduces other consequences of such a thermally stratified layer on the thermal evolution of Earth's core.

We further remark that, because the heat flux at the core-mantle boundary decreases at a small rate, the steady-flux assumption for the conductive temperature profile becomes more accurate. For example, the temperature difference between the PWSF scheme with $N = 1$ (equivalent to the steady-flux approach) and with $N = 7$ (or with any higher resolution) is below 0.3 degrees and the thickness of the thermally stratified region differs by less than 3 km.

[Place of table 1]

## 4.2 Mercury

[Place of figure 5]

In contrast to the silicate dominated nature of the Earth, planet Mercury consists largely (for ~70wt%) of core metal (Hauck et al., 2013). And while the thermal stratification in Earth's core is most likely a recent phenomenon, if it is present at all, the heat flux drops below the adiabatic heat flux at the core-mantle boundary of Mercury very early in its thermal evolution (e.g., Stevenson et al., 1983; Tosi et al., 2013; Knibbe and van Westrenen, 2018).

We first consider a thermal evolution scenario for Mercury similar to that developed to study thermal stratification in Earth's core (figure 4), in which we assumed that the temperature of Mercury's mantle is constant with time. That scenario (presented in Appendix C) reveals that the heat flux at the core mantle boundary is more than twice as high in thermal evolution runs that take the thermally stratified layer into account than in those runs that neglect thermal stratification (figure C1). The influence of thermal stratification in Mercury's core on the heat flux into the mantle is thus much larger than the influence of thermal stratification in Earth's core on the core-mantle boundary heat flux of about 3.3 % (figure 4). Because, additionally, the relative size of Mercury's mantle is small, we may expect that the thermal evolution of the mantle is substantially affected by thermal stratification in Mercury's core and provides additional feedback to the heat flux at the core-mantle boundary. For this reason, the influence of the thermal stratification on the temperature of the mantle should be incorporated for investigated the behaviour and implications of thermal stratification in the core of Mercury.

We couple the heat flux at the core-mantle boundary to a parametrized thermal evolution model that uses energy-balance principles to determine the evolution of the temperature of the convective mantle (e.g. Stevenson et al., 1983; Thiriet et al., 2019), which is described in Appendix B. Transport of heat from the core into the mantle, transport of heat from the mantle into the lithosphere, and the production of heat from radiogenic heat producing elements are considered in this energy-balance approach. We perform thermal evolution runs with the PWSF schemes with $N = 1$, $N = 4$ and $N = 8$. These choices for the resolution result in a similar effective resolution of the thermally stratified layer in Mercury's core that is about 2020 km in radius compared to the obtained resolution in the thermally stratified layer in Earth's core that is 3480 km in radius, for which we used $N = 1$, $N = 7$ and $N = 14$.

Results of the thermal evolution scenario for Mercury are presented in figure 5. The mantle temperature increases in the first ~800 million years by production of radiogenic heat (figure 5b), which is consistent thermal evolution models of Tosi et al. (2013) and Knibbe and van Westrenen (2018). Because of this increase in the mantle temperature, the thermal gradient in the core at the core-mantle boundary decreases rapidly in this period, from -6.8 $K \cdot m^{-1}$ at the onset of stratification (80 million years from the start) to -3.1·10$^{-4}$ K·m$^{-1}$ at the onset of core solidification (870 million years from the start). This time-variation of the thermal gradient at the core-mantle boundary is much smaller than in the example of figure 2, which thus represents a simplified scenario (simplified in the sense that the conductive domain is fixed in figure 2) where the boundary condition at the core-mantle boundary varies extremely rapidly with time. The smaller rate of decrease of the core-mantle boundary heat flux in this scenario for Mercury leads to radial variations in the cooling rate within the conductive region of maximally about 30 nW·m$^{-3}$, from about 0 nW·m$^{-3}$ at the core-mantle boundary to about 30 nW·m$^{-3}$ at the interface at ~1500 km radius (figure 5d). The PWSF schemes with $N = 4$ and $N = 8$, which effectively have three and six intervals for the stratified region, respectively, are practically indistinguishable (blue and red lines in figure 5). This indicates that these schemes closely represent the actual solution of the problem. The adequacy of these thermal evolution runs is further strengthened by a comparison with figure 2, which shows that radial variations in the cooling rate of about 117 nW·m$^{-3}$, which is a much larger variation in the cooling rate than the 30 nW·m$^{-3}$ radial variation in the cooling rate in the scenario studied here (figure 5d), are accurately represented by PWSF schemes with four intervals for the thermally stratified region.

[Place of figure 6]

The PWSF thermal evolution runs deviate substantially from the thermal evolution scheme that neglects thermal stratification (figures 4 and 5). In the first ~800 million years that the thermally stratified layer develops, the adiabatic region of the core cools at a much more rapid rate than the upper part of the core (figures 5b and 5d). Consequently, if thermal stratification is considered, the inner core starts to solidify about 1.4 billion years earlier and its present-day radius is about 500 km larger as compared to if thermal stratification is neglected (figure 5a). Such an earlier start of the inner core lengthens the time that a dynamo can be active in Mercury, for which the power largely originates from solidification of an inner core, according to this scenario. The temperature and heat flux at the core-mantle boundary are about 50 degrees higher (figure 5b) and almost doubled (figure 5c), respectively, compared to if thermal stratification is neglected. Because of the larger heat flux from the core to the mantle, the temperature of the mantle is about 23 degrees higher if thermal stratification is considered compared to if it is neglected (figure 5b). Of additional importance is that the transport of heat through the mantle is larger in the thermal evolution runs that consider thermal stratification, which is reflected by a larger heat flux into the lithosphere and a higher Rayleigh number (figure 6). The consideration of thermal stratification in the core is thus important for studying the vigour and duration of convection in Mercury's mantle.

Finally, we note that the PWSF scheme with $N = 1$ has a temperature error of maximally 3 degrees relative to the PWSF schemes with higher resolution, which occurs at the core-mantle boundary just prior to the onset of core solidification. After the onset of solidification of the inner core, radial variations in the cooling rate throughout the stratified region are diminished (figure 5d), such that the PWSF schemes with $N = 1$, $N = 4$ and $N = 8$ quickly become practically indistinguishable.

## 5 Summary

We presented a piece-wise steady-flux (PWSF) numerical scheme for solving one-dimensional conduction problems on a line segment (Appendix A) and in the radial direction of a spherically symmetric body or spherical shell (section 3). This numerical scheme adopts a 'steady-flux' solution of the conduction equation in each interval of the chosen discretization and assures that temperature is continuous and differentiable in the space domain. These smoothness properties are convenient for problems in which boundaries of the conductive domain vary with time. We applied the PWSF scheme in an energy-conserved approach for the thermal evolution of the cores of Earth and Mercury, with consideration of a conductive thermally stratified layer that develops below the core-mantle boundary when the heat flux drops below the adiabatic heat flux. The influence of a thermally stratified conductive region on the evolution of the planetary body is examined as well as the dependence of the model's accuracy to the applied resolution. The thermal coupling between the core and the mantle is also considered.

By considering thermal stratification in a planetary core, radial variations in the cooling rate are accounted for, whereas otherwise the distribution of energy in the core is fixed by the imposed adiabat. While the thermally stratified region develops below the core-mantle boundary, the deep part of the core cools more rapidly than the stratified region of the core. Therefore, the inner core grows to a larger size and the temperature and heat flux at the core-mantle boundary are higher if a thermally stratified region is considered. For the Earth, these implications are likely very minor and can be neglected in thermal evolution studies that are not specifically interested in the thermally stratified region itself. For Mercury, however, these implications are much larger and cannot be neglected. The age of the inner core can be underestimated by several billion years if thermal stratification is neglected, which has important consequences for Mercury's dynamo. The consideration of thermal stratification in Mercury's core is also important for the evolution of Mercury's mantle. It increases the mantle temperature, leads to a higher Rayleigh number and therefore a larger heat flux into the lithosphere and prolongation of mantle convection.

**Acknowledgements**

This project has received funding from the European Union's Horizon 2020 research and innovation program under the Marie Sklodowska-Curie grant agreement MERCURYREFINEMENT, with No 845354, awarded to JSK. Throughout the course of this project, JSK also received funding by a postdoctoral Marie-Curie Seal of Excellence fellowship (12Z622ON) of the FWO-Flanders. The authors also thank the Belgian Federal Science Policy Office (BELSPO) for the provision of financial support in the framework of the PRODEX Programme of the European Space Agency (ESA) under contract number PEA4000129360 and for the BRAIN-be2.0 project STEM.

## Data Availability

Upon final submission of this manuscript, we will make the Matlab code that is used to compute and plot the thermal evolution of scenario Mercury$^2$ (presented in figures 5 and 6) freely available at the online Zenodo repository (https://zenodo.org). The other thermal evolution scenarios and the test runs (figures 1-4) can be computed by minor changes in the parameter declaration of the code, as will be instructed by comments.

Vander Kaaden, K. E., F. M. McCubbin, A. A. Turner, and D. Kent Ross (2020), Constraints on the Abundances of Carbon and Silicon in Mercury's Core from experiments in the Fe-Si-C system, *Journal of Geophysical Research: Planets, 125*(5), e2019JE006239, doi: 10.1029/2019JE006239.

**Appendix A, The conduction problem on a line segment**

In this section, we develop the PWSF scheme for a conduction problem on a line segment

$$\frac{\partial T(t,x)}{\partial t} = \kappa \frac{\partial^2 T(t,x)}{\partial x^2}, \qquad x_0 \le x \le x_N, t > 0, \qquad (A1)$$

with $\kappa$ the thermal diffusivity. We discretized the domain $(x_0\ x_N)$ in $N$ intervals $(x_{i-1}\ x_i)$ with length $L_i = x_i - x_{i-1}$, for $i = 1, 2, \dots N$. It is convenient to introduce the array

$$U(t) = \begin{bmatrix} U_1(t) \\ U_2(t) \\ \vdots \\ U_N(t) \end{bmatrix} = \begin{bmatrix} \int_{x_0}^{x_1} T(t,x)dx \\ \int_{x_1}^{x_2} T(t,x)dx \\ \vdots \\ \int_{x_{N-1}}^{x_N} T(t,x)dx \end{bmatrix} \qquad (A2)$$

of dimension $N$, which is composed of the integrals of $T(t,x)$ over the $N$ subintervals. By integrating equation A1 with respect to $x$, we obtain

$$\frac{dU(t)}{dt} = \kappa \left( \begin{bmatrix} \frac{\partial T(t,x_1)}{\partial x} \\ \frac{\partial T(t,x_2)}{\partial x} \\ \vdots \\ \frac{\partial T(t,x_N)}{\partial x} \end{bmatrix} - \begin{bmatrix} \frac{\partial T(t,x_0)}{\partial x} \\ \frac{\partial T(t,x_1)}{\partial x} \\ \vdots \\ \frac{\partial T(t,x_{N-1})}{\partial x} \end{bmatrix} \right), \qquad (A3)$$

which we integrate with respect to time using forward finite differences. In equation (A3), the thermal gradients $\partial T(t,x_i)/\partial x$ for $i = 0$ and $i = N$ are given by the (Neumann) boundary conditions, whereas those for $i = 1, 2, \dots N-1$ need to be determined. A finite difference approximation to the $N-1$ unknown temperature gradients is

$$\frac{\partial T(t,x_i)}{\partial x} = \frac{2\left(T(t, x_i - \frac{1}{2}L_i) - T(t, x_i + \frac{1}{2}L_{i+1})\right)}{L_{i+1} + L_i}. \qquad (A4)$$

If $T$ is implicitly treated as a constant across each interval of constant size ($L_i = dx$ for $i = 1, 2, \dots N$), equations (A2-A4) yield

$$\frac{T(t+dt,x) - T(t,x)}{dt} = \kappa \frac{T\left(t, x_i - \frac{1}{2}dx\right) - 2T\left(t, x_i + \frac{1}{2}dx\right) + T\left(t, x_{i+1} + \frac{1}{2}dx\right)}{dx^2}, \quad x_i < x < x_{i+1} \quad (A5)$$

which is a finite volume numerical scheme for the conduction equation (equation A1) that uses first-order forward finite differences with respect to time and central second-order differences with respect to $x$, also known as the method of lines. Instead of using equation (A4) for estimating the temperature gradients at the interval boundaries, the piece-wise steady-flux (PWSF) numerical scheme is developed by assuming a steady flux form for $T(t,x)$ in each interval

$$T_i(t,x) = -\frac{S_i(t)}{2\kappa}x^2 + A_i(t)x + B_i(t), \;\; x_{i-1} < x < x_i, i = 1,2,\dots,N, \quad (A6)$$

with

$$S_i(t) = \kappa\frac{\frac{\partial T(t,x_{i-1})}{\partial x} - \frac{\partial T(t,x_i)}{\partial x}}{L_i}, \qquad (A7)$$

$$A_i(t) = \frac{\partial T(t,x_{i-1})}{\partial x} + \frac{S_i(t)}{\kappa}x_{i-1}, \qquad (A8)$$

and

$$B_i(t) = \frac{U_i + \frac{1}{6\kappa}S_i(t)({x_i}^3 - {x_{i-1}}^3) - \frac{1}{2}A_i(t)({x_i}^2 - {x_{i-1}}^2)}{L_i}. \qquad (A9)$$

The gradients $\partial T(t,x_i)/\partial x$ are determined by imposing continuity of $T$ at each interval boundary (i.e., $T_i(t,x_i) = T_{i+1}(t,x_i)$, for $i = 1,2,\dots\ N-1$). The $N-1$ continuity conditions can be written as

$$My = b, \qquad (A10)$$

with

$$y = \begin{bmatrix} \frac{\partial T(t,x_1)}{\partial x} \\ \frac{\partial T(t,x_2)}{\partial x} \\ \vdots \\ \frac{\partial T(t,x_{N-1})}{\partial x} \end{bmatrix}, \qquad (A11)$$

$$b = \begin{bmatrix} \frac{U_2(t)}{L_2} - \frac{U_1(t)}{L_1} + \frac{\partial T(t,x_0)}{\partial x}\left(\frac{2{x_1}^2 - 2x_0x_1 - {x_0}^2}{2L_1} - x_1 + \frac{3x_0({x_1}^2 - {x_0}^2) - ({x_1}^3 - {x_0}^3)}{6{L_1}^2}\right) \\ \frac{U_3(t)}{L_3} - \frac{U_2(t)}{L_2} \\ \vdots \\ \frac{U_{N-1}(t)}{L_{N-1}} - \frac{U_{N-2}(t)}{L_{N-2}} \\ \frac{U_N(t)}{L_N} - \frac{U_{N-1}(t)}{L_{N-1}} + \frac{\partial T(t,x_N)}{\partial x}\left(\frac{3x_{N-1}({x_N}^2 - {x_{N-1}}^2) - ({x_N}^3 - {x_{N-1}}^3)}{6{L_N}^2} - \frac{{x_{N-1}}^2}{2L_N}\right) \end{bmatrix}, (A12)$$

and a tridiagonal matrix $M$ with

$$M_{i,i-1} = x_i + \frac{2x_{i-1}x_i - 2{x_i}^2 + {x_{i-1}}^2}{2L_i} + \frac{({x_i}^3 - {x_{i-1}}^3) - 3x_{i-1}({x_i}^2 - {x_{i-1}}^2)}{6{L_i}^2};\ 2,3,\dots,N-1, \quad (A13)$$

$$M_{i,i} = \frac{{x_i}^2 - 2x_{i-1}x_i}{2L_i} + \frac{3x_{i-1}({x_i}^2 - {x_{i-1}}^2) - ({x_i}^3 - {x_{i-1}}^3)}{6{L_i}^2} - x_i + \frac{{x_{i+1}}^2 - 2{x_i}^2}{2L_{i+1}} + \frac{3x_i({x_{i+1}}^2 - {x_i}^2) - ({x_{i+1}}^3 - {x_i}^3)}{6{L_{i+1}}^2}; i = 1,2,\dots,N-1, \quad (A14)$$

and

$$M_{i,i+1} = \frac{{x_i}^2}{2L_{i+1}} + \frac{({x_{i+1}}^3 - {x_i}^3) - 3x_i({x_{i+1}}^2 - {x_i}^2)}{6{L_{i+1}}^2};\ i = 1,2,\dots,N-2. \quad (A15)$$

The $N-1$ thermal gradients $\partial T(t,x_i)/\partial x$ are solved from equation (A10) from knowledge of $U$ at time $t$ using standard linear algebra and used to integrate $U$ forward in time by equation (A3). The temperature profile can be reconstructed by equations (A6-A9).

[Place of figure A1]

We performed forward integrations of the PWSF numerical scheme with $N = 16$, $N = 4$ and $N = 1$ on a line segment $x_0 < x < x_N$. Without loss of generality, we set $x_0 = 0$. We choose inhomogeneous Neumann boundary conditions that are constant over time

$$\frac{\partial T(t,x_0)}{\partial x} = g_0, \qquad \frac{\partial T(t,x_N)}{\partial x} = g_N. \qquad (A16)$$

With an initial condition $T(0,x)$, the analytical solution of this problem is given by

$$T(t,x) = \frac{g_N - g_0}{2x_N}x^2 + g_0 x + \frac{g_N - g_0}{x_N}\kappa t + \frac{a_0}{2} + \sum_{n=1}^{\infty} a_n\, e^{-\kappa t\left(\frac{n\pi}{x_N}\right)^2} \cos\left(\frac{n\pi x}{x_N}\right), \quad (A17)$$

where

$$a_n = \frac{2}{x_N}\int_0^{x_N}\left(T(0,x) - \frac{g_N - g_0}{2x_N}x^2 - g_0 x\right)\cos\left(\frac{n\pi x}{x_N}\right)\mathrm{dx}. \qquad (A18)$$

We use boundary conditions $g_0 = 0$, $g_N \neq 0$ a constant, and initial condition $T(0,x) = 0$ (figure A1). Because a thermal gradient is imposed at $x_N$, time-variations of temperature (the rate of heating or cooling) propagates from $x_N$ towards $x_0$. We recall that the conduction equation (A1) relates the heating (or cooling) rate to the Laplacian (in this linear case equal to the second derivative) of the temperature profile, which for the PWSF numerical scheme is piecewise constant over each interval (equal to $-S$). Additionally, the PWSF scheme ensures conservation of energy on individual intervals by equation (A3). For these reasons, the PWSF numerical scheme approximates the solution of the conduction equation with high accuracy if the adopted intervals are smaller than the spatial scale of variations in the heating (or cooling) rate. For example, at $t = \tau/4^5$, with $\tau = (x_N - x_0)^2/\kappa$ the conduction timescale, the heating (or cooling) rate is significant only in the small outermost region of the sphere with $x/x_N > 0.9$ (figure 1a). Differences between the analytical solution of the conduction equation and the temperature profile of the PWSF schemes with $N = 4$ and $N = 1$ are, therefore, significantly in error whereas the errors of the PWSF scheme with $N = 16$ are small, at $t = \tau/4^5$ (figure A1a). As time progresses with the constant boundary condition, the heating (or cooling) rate distributes across the conductive domain and the piece-wise constant Laplacian of temperature of the low resolution PWSF schemes become more appropriate. The differences between the analytical solution and the temperature profile of the PWSF scheme $N = 4$ are small since $t \sim \tau/4^3$ (figure 1b). Since $t \sim \tau/4$, radial variations in the cooling or heating rate across the conductive domain have practically disappeared and also the PWSF scheme with $N = 1$ accurately describes the temperature profile and the thermal evolution (figure 1c).

**Appendix B**

To parametrize the core of the Earth, we largely follow the scenario of Labrosse et al. (1997). The adiabatic parameters $T_2$ and $T_3$ (listed in table 1 of the main paper) are fitted to the adiabat that is used in Labrosse et al. (1997). The density of the core is assumed constant, such that the pressure profile is

$$P(r) = P_c + \rho^2 \frac{2\pi}{3} G\left({R_c}^2 - r^2\right), \qquad \text{(B1)}$$

with $P_c$ the pressure at the core-mantle boundary and $G$ the gravitational constant. The solidification temperature is assumed to increase linearly with increasing pressure, and takes a value of 5000 K at a radius of 1221 km and 5270 K at the Earth's center (Labrosse et al., 1997). The rate of energy production due to inner core growth is given by

$$Q(t) = 4\pi {R_i}^2 \left(L + E_g\right) \rho \frac{dR_i}{dt}, \qquad \text{(B2)}$$

where $L$ and $E_g$ denote the latent and gravitational energy production per unit mass of solidified core material. The latent heat production per unit mass is given by

$$L = T(R_i)\Delta S, \qquad \text{(B3)}$$

with $\Delta S$ the change in entropy by solidification per unit mass. The production of gravitational energy per unit mass is parametrized as

$$E_g = \Delta\rho \frac{4\pi}{3} G {R_c}^2 \left( \frac{3}{10} - \frac{1}{2} \left( \frac{R_i}{R_c} \right)^2 \right) \qquad \text{(B4)}$$

with $\Delta\rho$ the density jump upon solidification. The imposed heat flux at the core mantle boundary decreases linearly with time from 75 mW·m$^{-2}$ (equal to 11.41 TW of surface integrated flux at $R_c$) at 4.5 Ga to 25.3 mW·m$^{-2}$ (3.85 TW) at present.

In figure 4 of the main paper, the heat flux at the core-mantle boundary is determined from the thermal gradient in the lower thermal boundary layer in the mantle

$$F_c = k_m \frac{T_c - T_m}{\delta}, \qquad \text{(B5)}$$

with $k_m$ the thermal conductivity of the mantle, $T_m$ the temperature at the base of the convective part of the mantle, and $\delta$ the thickness of the mantle's lower thermal boundary layer. We follow a parametrization of $\delta$ that is commonly used in parametrized models for convection in planetary mantles (e.g., Stevenson et al., 1983; Thiriet et al., 2019)

$$\delta^3 = \frac{Ra_{cr} v K}{g_m \alpha \Delta T}, \qquad \text{(B6)}$$

where $Ra_{cr}$ is the critical Rayleigh number that is set at 2000 for the lower thermal boundary layer of the mantle (Stevenson et al., 1983), $v$ is the average kinematic viscosity of the thermal boundary layer, $K$ is the thermal diffusivity of the mantle, $g_m$ is the gravity in the mantle, $\alpha$ is the thermal expansion of the mantle, and $\Delta T$ is a temperature difference that, for the lower thermal boundary layer of interest here, equals $T_c - T_m$. The kinematic viscosity is treated as a temperature dependent variable

$$v(T) = v_0 \exp\left(\frac{A}{T}\right), \qquad \text{(B7)}$$

where $v_0$ and $A$ are constants (e.g., Stevenson et al., 1983), and is for the lower thermal boundary layer of the mantle evaluated at $T = (T_c - T_m)/2$. The kinematic viscosity in Earth's lower mantle is predicted to vary from ~ $10^{19}$ m$^2$·s$^{-1}$ at the base of the convecting mantle to ~ $10^{16}$ m$^2$·s$^{-1}$ near the core-mantle boundary (e.g., Mitrovica and Forte, 2004; Steinberger and Calderwood, 2006; Ciskova et al., 2012; e.g. figure 1 of Ciskova et al., 2012). We choose values of $v_0$ and $A$ that lead to a kinematic viscosity of the lower thermal boundary layer in the middle range of these values. The initial core-mantle boundary temperature is set at 4300 K, instead of the value of 4247 K when the scenario of Labrosse et al. (1997) is reproduced, to obtain a similar present-day radius of the inner core.

We initially follow the same procedure for studying the thermal evolution of Mercury. We use core properties of Mercury similar to those used by Knibbe and van Westrenen (2018), with a core density for Mercury of ~7200 kg·m$^{-3}$, thermal conductivity of 40 W·m$^{-1}$·K$^{-1}$, and heat capacity of 840 J·kg$^{-1}$·K$^{-1}$. Considering a thermal expansion between 5.5·10$^{-5}$ K$^{-1}$ and 9.5·10$^{-5}$ K$^{-1}$ at the ~5 GPa core-mantle boundary pressure of Mercury (Secco, 2017), the adiabatic gradient at Mercury's core-mantle boundary of ~2020 km radius is somewhere between -5.5·10$^{-4}$ K·m$^{-1}$ and -9.5·10$^{-4}$ K·m$^{-1}$. We set $T_2$ equal to -2·10$^{-14}$ K·m$^{-2}$ and $T_3$ to -1.5·10$^{-20}$ K·m$^{-3}$, which results in an adiabatic gradient at Mercury's core-mantle boundary of -7·10$^{-4}$ K·m$^{-1}$. The solidification curve is taken from Morard et al. (2011)

$$T_S = T_{S,0}\left(\frac{P}{a}+1\right)^{1/c}, \qquad \text{(B8)}$$

with $T_{S,0} = 1510$ K, $a = 10$ GPa and $c = 3$. The parameter $T_{S,0}$ is increased by 32 K compared to the value reported by Morard et al. (2011) to increase the solidification temperature at 21 GPa by ~45 K compared to the solidification curve of Fe-15wt%Si-5wt%Ni. Silicon is expected to lower the solidification temperature by ~15K per wt% Si at 21 GPa (Kuwayama and Hirose, 2004). The solidification curve that is adopted here is thus representative for a core composition of Fe-12wt%Si-5wt%Ni. Such a composition is consistent with geodetic and geochemical constraints of Mercury (Vander Kaadden et al., 2020; Steenstra and van Westrenen, 2020; Knibbe et al., 2021), although such a core composition for Mercury may lead to a larger inner core compared to that inferred from magnetic field simulations (Knibbe et al., 2021; and references therein). We assume that the combined latent heat and gravitational heat that is released by core solidification $\left(L + E_g\right)$ amounts to 5·10$^5$ J·kg$^{-1}$. The temperature in the lower mantle $(T_m)$ is set to 1800 K, which is in the range of present-day mantle temperatures of the thermal evolution scenarios presented by various thermal evolution studies for Mercury (Michel et al., 2013; Tosi et al., 2013; Knibbe and van Westrenen et al. 2018). The results of this scenario for Mercury are presented in appendix C.

In figures 5 and 6 of the main paper, the influence of thermal stratification in Mercury's large core on the evolution of the mantle is studied by a implementing a parametrization of the convective mantle. The rate of temperature change in the lower mantle is governed by an energy balance equation of the convective part of the mantle

$$\frac{dT_m}{dt} = \frac{Q_m}{c_m} + \frac{4\pi\left(F_c {R_c}^2 - F_l {R_l}^2\right)}{\frac{4\pi}{3}\left({R_l}^3 - {R_c}^3\right)\rho_m c_m}, \qquad \text{(B9)}$$

where $R_l$ is the radius at the base of the lithosphere, $F_l$ is the heat flux from the convective mantle into the lithosphere, $c_m$ and $\rho_m$ are the heat capacity and density of Mercury's mantle, respectively, and $Q_m$ is the energy production of the mantle per unit mass. The thickness of the lower thermal boundary layer of the mantle and the heat flux at the core mantle boundary are given by equations (B5-B7). The heat flux into the lithosphere is given by

$$F_l = k_m \frac{T_m - T_l}{\delta_l}, \qquad \text{(B10)}$$

with $T_l$ the temperature at the base of the lithosphere, and $\delta_l$ the thickness of the upper thermal boundary layer of the convecting mantle. The temperature at the base of the lithosphere is estimated by

$$T_l = T_m - a_{rh} \frac{{T_m}^2}{A}, \qquad \text{(B11)}$$

with $a_{rh}$ set at 2.54 (Thiriet et al., 2019). The thickness $(\delta_l)$ of the upper thermal boundary layer is calculated by the same equation as the thickness of the lower thermal boundary layer (equation B6), with the viscosity evaluated at temperature $T_m$, instead of evaluating viscosity at $(T_c - T_m)/2$ as is done for the lower thermal boundary layer. It is well established that the critical Rayleigh number for the upper thermal boundary layer of the convecting mantle should be set lower than that for the lower thermal boundary layer (e.g., Stevenson et al., 1983). Accordingly, we set $Ra_{cr}$ at 500 in equation (46) for calculating $\delta_l$, instead of the value of 2000 that is used for calculating the thickness of the lower thermal boundary layer. Heat transfer through the lithosphere and through the upper thermal boundary layer occurs by conduction. The conduction timescale of a silicate layer of 100 km or 300 km thick, which are feasible sizes of the combined lithosphere and upper thermal boundary layer, ranges between 300 million years and 3 billion years. We have seen that the conductive temperature profile converges to a steady flux form, which is close to linear in radius in the lithosphere, in about one-fourth of the conduction timescale (figure A1 of Appendix A and figures 1 and 2 of the main paper). Because this convergence is relatively fast in comparison to the considered geological timescale of multiple billion years, the thermal gradients in the lithosphere and the upper thermal boundary layer are expected to be very similar (equation 18 of Thiriet et al., 2019). Based on these considerations, the thickness of the lithosphere is approximately given by

$$\left(R_p - R_l\right) \approx \delta_l \frac{T_l - T_p}{T_m - T_l}, \qquad \text{(B12)}$$

with $T_p = 440$ K the temperature at Mercury's surface. In test runs of this case study, the right-hand side varies between 140 km and 260 km throughout the thermal evolution. We, therefore, assume a lithosphere of 200 km thick, and accordingly set $R_l$ at 2240 km throughout the evolution. Time variations of the thickness of the lithosphere are thus neglected, for sake of simplicity.

Isotopes of thorium, uranium and potassium contribute to the radiogenic heating of Mercury's mantle, each with its own half-life and initial heat production. We describe their combined contribution as a single variable that decays exponentially with time

$$Q_m = Q_0 \exp\left(\frac{-t}{\tau_{1/2}} \cdot \ln(2)\right), \qquad \text{(B13)}$$

with $Q_0$ the radiogenic heat production rate at the beginning of the evolution and $\tau_{1/2}$ the half-life time. The total present-day radiogenic energy production in the mantle is estimated between $1.88 \cdot 10^{-12}$ W·kg$^{-1}$ and $4.55 \cdot 10^{-12}$ W·kg$^{-1}$, whereas the initial heat production is estimated between $2.3 \cdot 10^{-11}$ W·kg$^{-1}$ and $5.4 \cdot 10^{-12}$ W·kg$^{-1}$ (Padovan et al., 2015). We set $Q_0$ at $1.5 \cdot 10^{-11}$ W·kg$^{-1}$ and set the half-life at 1.9 billion years to obtain a present-day radiogenic heat production near the middle of the estimated values.

All adopted parameters for the four case studies are assembled in table 1 of the main paper.

**Appendix C Results of case study 'Mercury1'**

[Place of figure C1]

Thermal evolution runs for Mercury1, where the heat flux is controlled by the thermal gradient in the lower mantle that has a constant temperature, are shown in figure C1. The heat flux at the core-mantle boundary drops below the adiabatic heat flux at 137 million years after the start of the evolution, such that a thermally stratified region develops below the core-mantle boundary. The cooling rate in the thermally stratified region varies significantly with radius, particularly in the time period before the inner core has started to precipitate. The temperature at the core-mantle boundary dropped 69 degrees between 137 million year and 1 billion years, whereas the temperature at 1500 km radius dropped 154 K in these 863 million years. Averaged over this time period, the secular cooling ($\Delta T c \rho$ / $\Delta t$) equals 15.3 nW·m$^{-3}$ at the core-mantle boundary and 34.1 nW·m$^{-3}$ at the 1500 km radius. Correspondingly, the difference in parameter $S$ between the uppermost and lowermost part of the conductive region is about 20 nW·m$^{-3}$, of which figure 2 of the main paper shows that such variations in the cooling rate can be considered appropriately by using only 2 or four intervals for the thermally stratified region. The PWSF schemes with $N = 4$ and $N = 8$ effectively discretise the thermally stratified region into three and six intervals (figure D1d), such that these thermal evolution runs both closely approximate the actual solution to the problem.

Since the onset of thermal stratification, the thermal evolution runs that consider thermal stratification facilitate more transport of heat from the adiabatic part of the core into the uppermost part of the outer core compared to if thermal stratification is neglected. As a result, the deeper part of the core cools at a more rapid rate and the core-mantle boundary cools at a reduced rate in the numerical schemes that consider thermal stratification, compared to when it is neglected. The solidification of the inner core starts at about 1 billion years in the thermal evolution runs that consider stratification and at about 1.5 billion years in the thermal evolution run that neglects thermal stratification. The excessive radial transport of heat in the thermal evolution runs that consider stratification leads to about 60 degrees higher temperature at the core-mantle boundary at present day, and more than doubles the heat flux into the mantle, compared to if thermal stratification is neglected. However, the temperature of the mantle is set constant in this case study. In reality, the thermal evolution of the mantle is sensitive to the heat flux at Mercury's core mantle boundary, particularly because the mantle of Mercury is less than half the size of Mercury's core. Therefore, a more elaborate description of the mantle's thermal behaviour is needed to study the influence of thermal stratification in Mercury's core. This is done in figures 5 and 6 of the main paper, using the method that is described in Appendix B.

**Figures and tables**

Figure 1

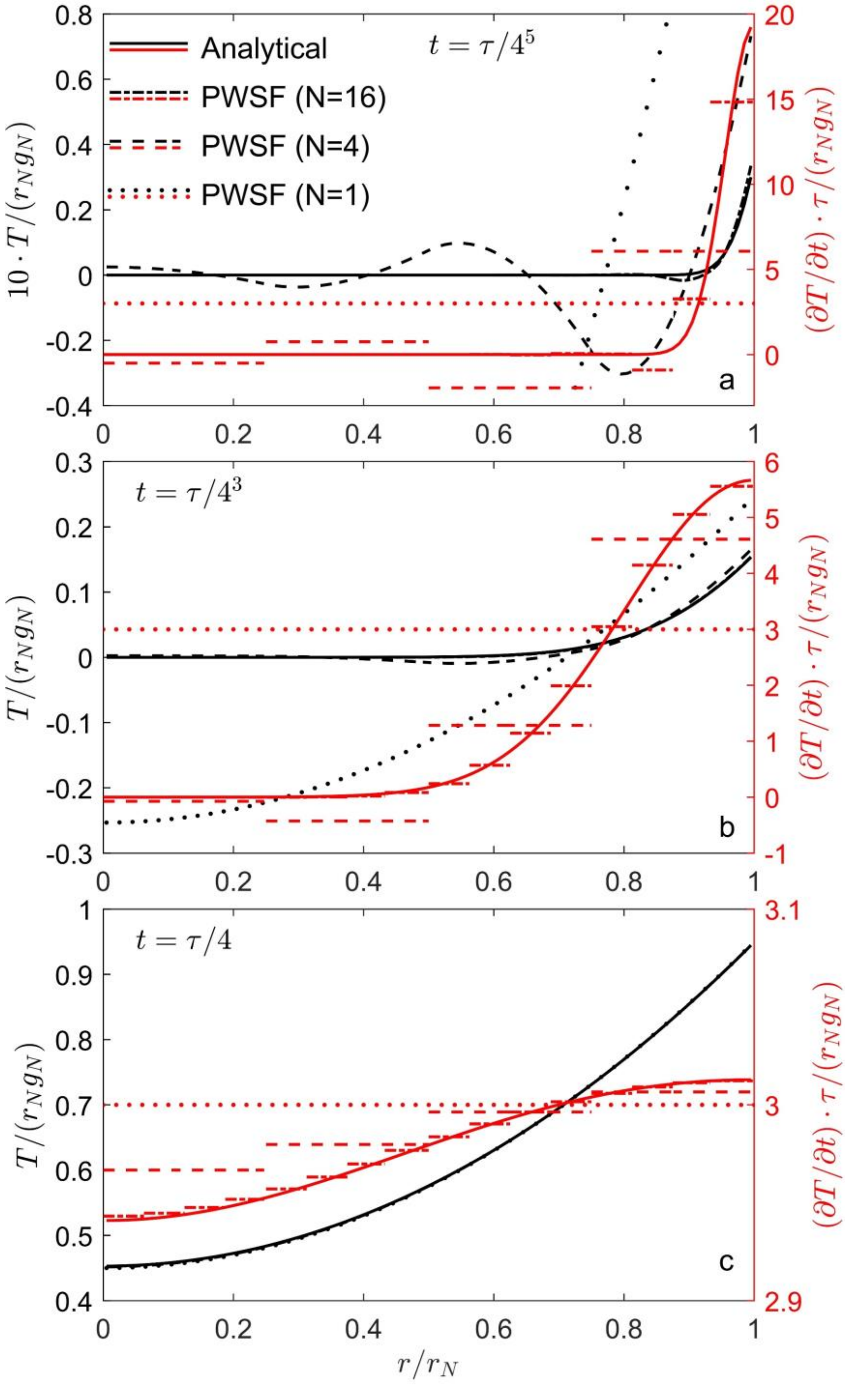

Caption of figure 1: Comparison of the piece-wise steady flux (PWSF) numerical scheme with one interval ($N = 1$, dotted), four intervals ($N = 4$, dashed), or sixteen intervals ($N = 16$, dashed-dotted), and the analytical solution (solid) for conduction though a sphere of radius $r_N$. Inhomogeneous Neumann boundary conditions are adopted with thermal gradients $\partial T(t, r_0)/\partial r = 0$ and $\partial T(t, r_N)/\partial r = g_N$. The diffusion timescale is $\tau = {r_N}^2/\kappa$. Panels (a), (b), and (c), show the non-dimensional temperature $\kappa T(r)/(r_N g_N)$ (black, left Y-axis) at $t = \tau/4^5$, at $t = \tau/4^3$, and at $t = \tau/4$, respectively. In red, the non-dimensional cooling rate $\left((\partial T/\partial t) \cdot \tau/(r_N g_N)\right)$ of the analytical solution is plotted (red solid) and compared to the non-dimensional secular cooling parameters of the temperature profile $(-S(r) \cdot \tau/(r_N g_N))$ of the PWSF numerical schemes, which are constant over each interval. Temperature of panel (a) is magnified relative to panels (b) and (c), to visually discriminate the temperature profile of the PWSF scheme with $N = 16$ from the analytical solution.

Figure 2

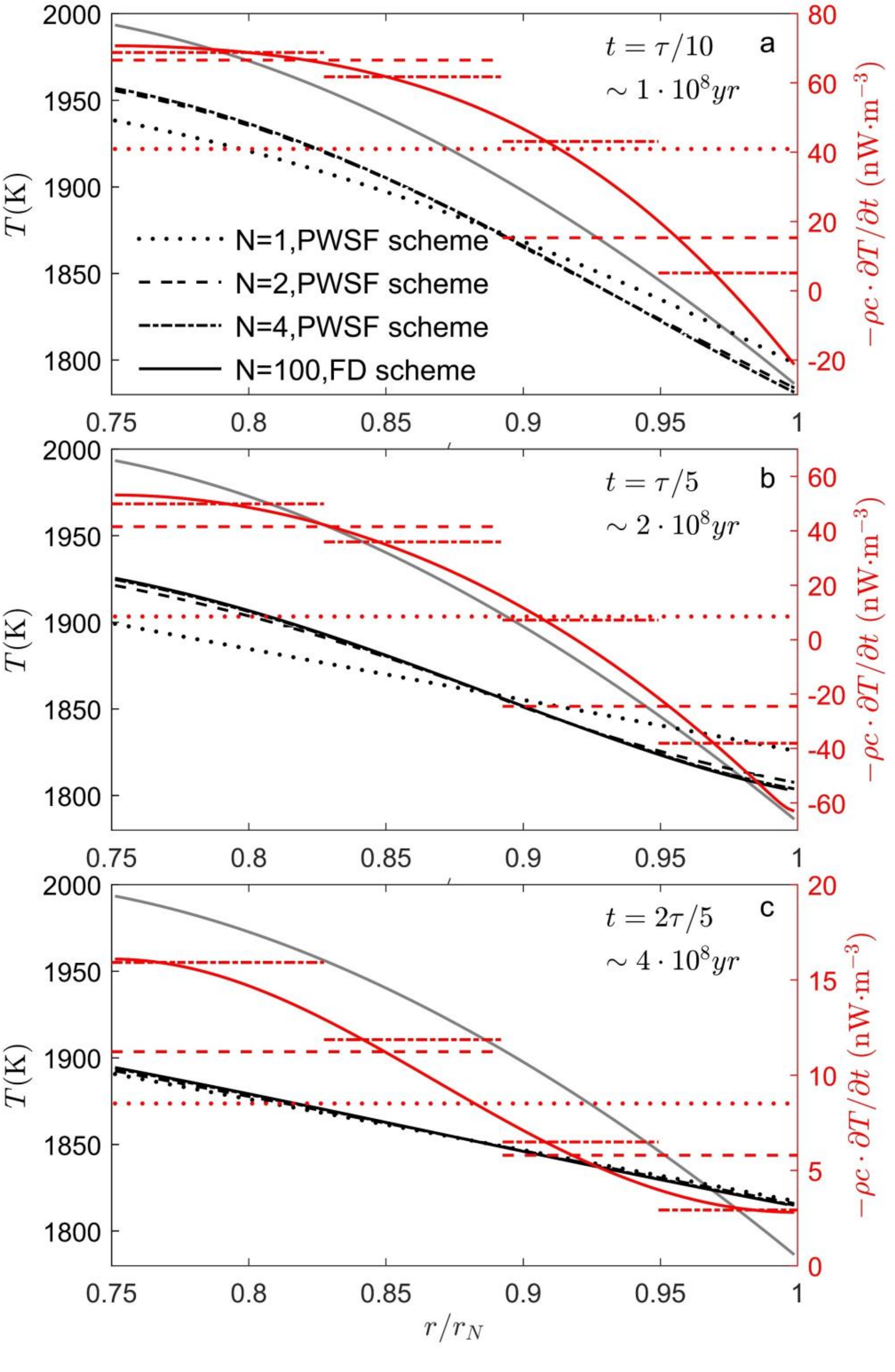

Caption of figure 2: Comparison of conduction across a spherical shell bounded by $r_0 = 1.5 \cdot 10^6$m and $r_N = 2 \cdot 10^6$m using the piece-wise steady flux (PWSF) numerical scheme ($N = 1$ dotted, $N = 2$ dashed, and $N = 4$ dashed dotted). The density is set at 7200 kg·m$^{-2}$, heat capacity at 840 J·kg$^{-1}$·K$^{-1}$ and thermal conductivity at 50 W·m$^{-1}$·K$^{-1}$. The benchmark numerical scheme is produced by estimating the thermal gradients at the interval boundaries by finite differences (FD) (equation 21), with $N = 100$ (solid). Inhomogeneous Neumann boundary conditions are adopted, with the gradient boundary condition $\partial T(t, r_N)/\partial r = g_N$ decreasing linearly with time from $-6.5 \cdot 10^{-4}$ K·m$^{-1}$ at $t = 0$ to $-1.5 \cdot 10^{-4}$ K·m$^{-1}$ at $t \geq \tau/5$. The gradient boundary condition $\partial T(t, r_0)/\partial r = g_0$ is set at $-1.5 \cdot 10^{-4}$ K·m$^{-1}$, constant over time. The initial thermal profile (solid grey) is assumed of steady-flux form (equation 12), derived by the initial boundary conditions and a volumetric average temperature of the conductive domain of 1900 K. Panels (a), (b) and (c) show the obtained results at $t = \tau/10$, $t = \tau/5$ and $t = 2\tau/5$, respectively. Temperature is plotted in black. The local cooling rate of the benchmark model is plotted in solid red. The average secular cooling of the PWSF scheme of each interval (the $S_i$ for $i = 1, 2, \dots, N$) are also plotted in red. The temperature profile of the PWSF schemes with $N = 2$ (dashed) and particularly with $N = 4$ (dashed dotted) are largely overplotted by the benchmark model. Intervals are taken of equal volume rather than equal thickness (which was done in figure 1).

Figure 3

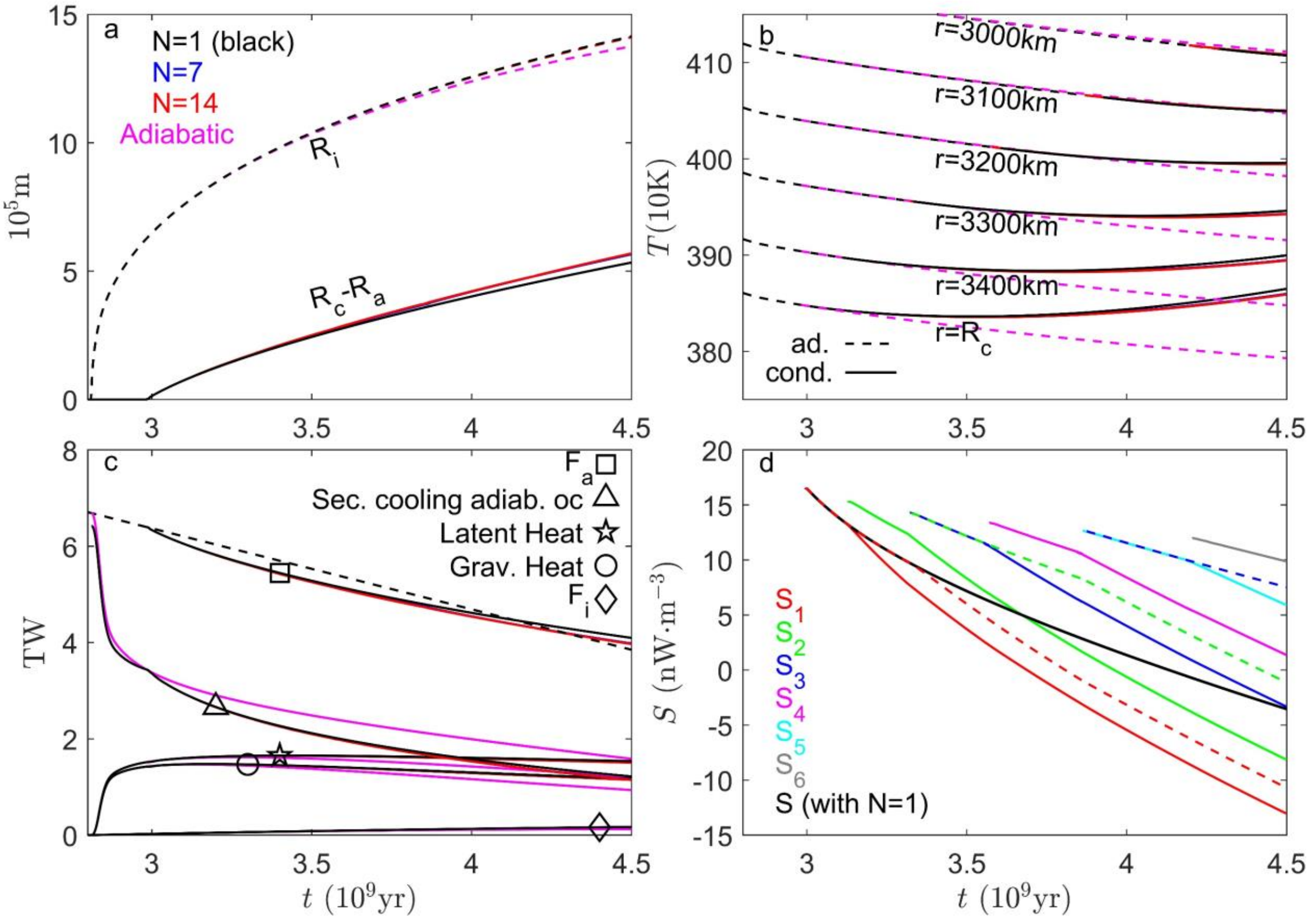

Caption of figure 3: Thermal evolution of Earth's core using parameters of Labrosse et al. (1997) (see scenario Earth[1] in table 1 for the adopted parameters). (a) The radius of the inner core $(R_i)$ and the thickness of the conductive layer $(R_c - R_a)$. (b) Temperature at 3000 km, 3100 km, 3200 km, 3300 km, 3400 km radius, and at the core-mantle boundary $(R_c)$ of 3480 km. The dashed and solid lines indicate the adiabatic and conductive temperature profile, respectively. (c) The energy balance of the adiabatic region of the liquid outer core. The heat that is transported across the interface $(R_a)$ is denoted by $F_a$ (square symbol). The secular cooling in the adiabatic part of the liquid core is denoted by the triangle symbol. The release of latent heat and of gravitational energy upon solidification are denoted by pentagram and circle symbols, respectively. The heat flux through the inner core boundary, which equals the secular cooling of the solid inner core, is denoted by diamond symbol. In panels (a), (b), and (c), the thermal evolutions computed with $N = 1$, $N = 7$, and $N = 14$ are plotted in black, blue, and red, respectively. Magenta lines correspond to a thermal evolution run where thermal stratification is not considered and the core's profile is considered to be adiabatic for the entire evolution. Blue lines are almost completely overplotted by red lines. Red lines are largely overplotted by black lines. (d) The parameter $S$ in the thermal evolution with $N = 1$ is plotted in black. The $S_i$ for $i$ from 1 to 6, plotted by solid-coloured lines, are the parameters of the temperature profile in the six uppermost conductive shells in the thermal evolution produced with $N = 14$. The $S_i$ for $i$ from 1 to 3, plotted by dashed-coloured lines, are the parameters of the temperature profile across the three uppermost conductive shells in the thermal evolution produced with $N = 7$.

Figure 4

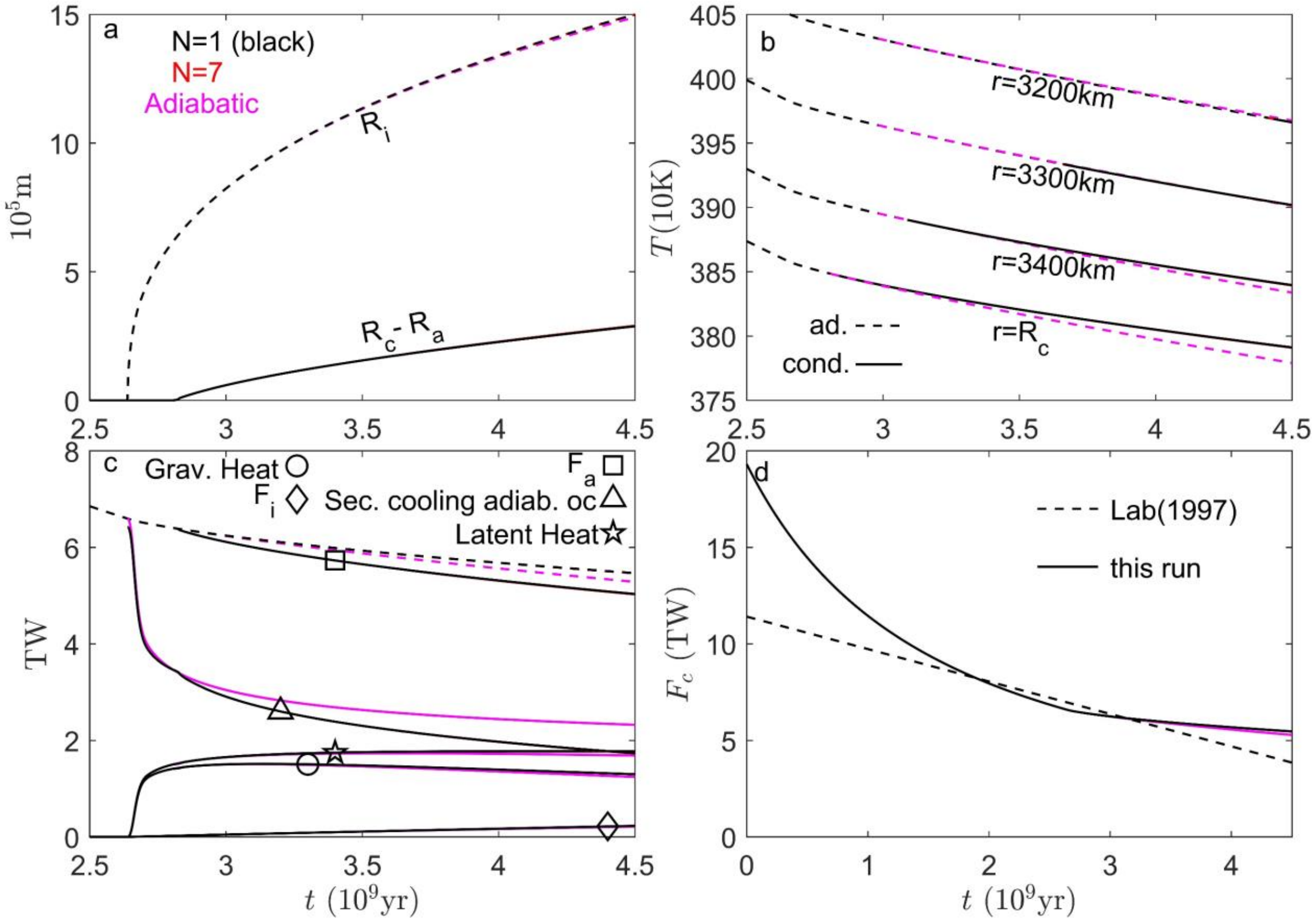

Caption of figure 4: Thermal evolution of Earth's core, coupled to the thermal boundary layer of the mantle (see scenario Earth[2] in table 1 for the adopted parameters). (a) The radius of the inner core ($R_i$) and the thickness of the conductive layer ($R_c - R_a$) set out against time. (b) Temperature at 3200 km, 3300 km, 3400 km radius, and at the core-mantle boundary ($R_c$) of 3480 km. The dashed and solid lines indicate the adiabatic and conductive temperature profile, respectively. (c) The energy balance of the adiabatic region of the liquid outer core. The heat that is transported across the interface ($R_a$) is denoted by $F_a$ (square symbol). The secular cooling in the adiabatic part of the liquid core is denoted by the triangle symbol. The release of latent heat and of gravitational energy upon solidification are denoted by pentagram and circle symbols, respectively. The heat flux through the inner core boundary, which equals the secular cooling of the solid inner core, is denoted by diamond symbol. (d) The heat flux at the core-mantle boundary of this run (solid line) and of the case reproducing Labrosse et al. (1997) that is presented in figure 3 (dashed line). (general) Thermal evolution runs with $N = 1$ and $N = 7$ are plotted in black and red, respectively. Red lines are almost completely overplotted by black lines. Magenta lines correspond to a thermal evolution run where the core's profile is considered to be adiabatic for the entire evolution and thermal stratification is, thus, not considered.

Figure 5

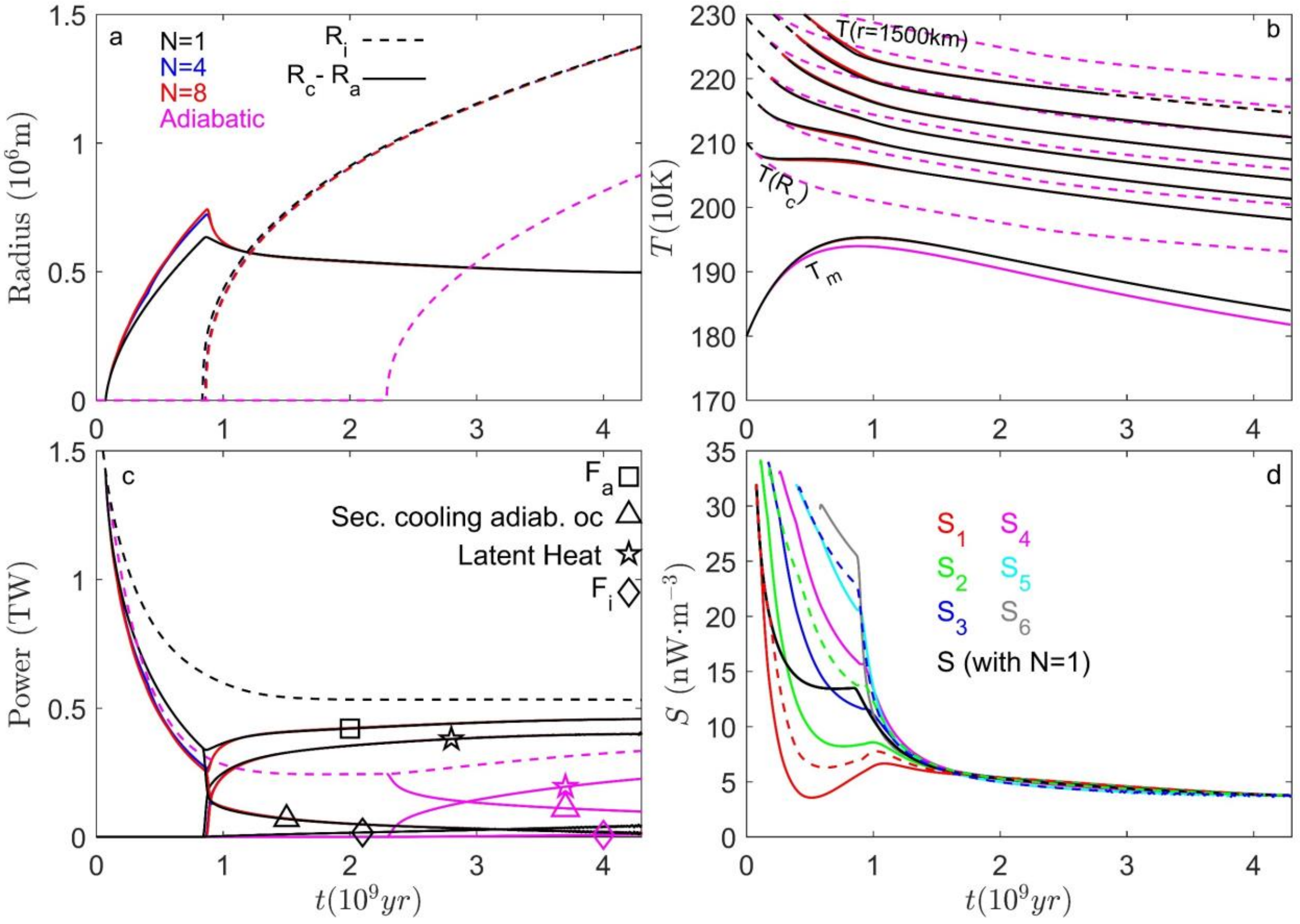

Caption of figure 5: Thermal evolution of Mercury's core, coupled to the thermal boundary layer of the mantle (see scenario Mercury[2] in table 1 for the adopted parameters). (a) The radius of the inner core $(R_i)$ and the thickness of the conductive layer $(R_c - R_a)$ set out against time. (b) Temperature at a radius of 1500 km, 1600 km, 1700 km, 1800 km, 1900 km radius, and at the core-mantle boundary $(R_c)$ of 2020 km. The dashed and solid lines indicate the adiabatic and conductive temperature profiles, respectively. (c) The energy balance of the adiabatic region of the liquid outer core. The heat that is transported across the interface $(R_a)$ is denoted by $F_a$ (square symbol). The secular cooling in the adiabatic part of the liquid core is denoted by the triangle symbol. The release of latent (and gravitational) energy upon solidification is denoted by pentagram symbol. The heat flux through the inner core boundary, which equals the secular cooling of the solid inner core, is denoted by diamond symbol. (Panels a, b, and c) Thermal evolution runs with $N = 1$, $N = 4$ and $N = 8$ are plotted in black, blue, and red, respectively. Blue lines are almost completely overplotted by red lines, whereas red lines are largely overplotted by black lines. Magenta lines correspond to a thermal evolution run where thermal stratification is not considered and the core's profile is considered to be adiabatic for the entire evolution. (d) The parameter $S$ in the thermal evolution with $N = 1$ is plotted in black. The $S_i$ for $i$ from 1 to 6, plotted by solid-coloured lines, are the parameters of the temperature profile across the six uppermost conductive intervals in the thermal evolution produced with $N = 8$. The $S_i$ for $i$ from 1 to 3, plotted by dashed-coloured lines, are the corresponding parameters of the temperature profile across the three uppermost conductive intervals in the thermal evolution produced with $N = 4$.

Figure 6

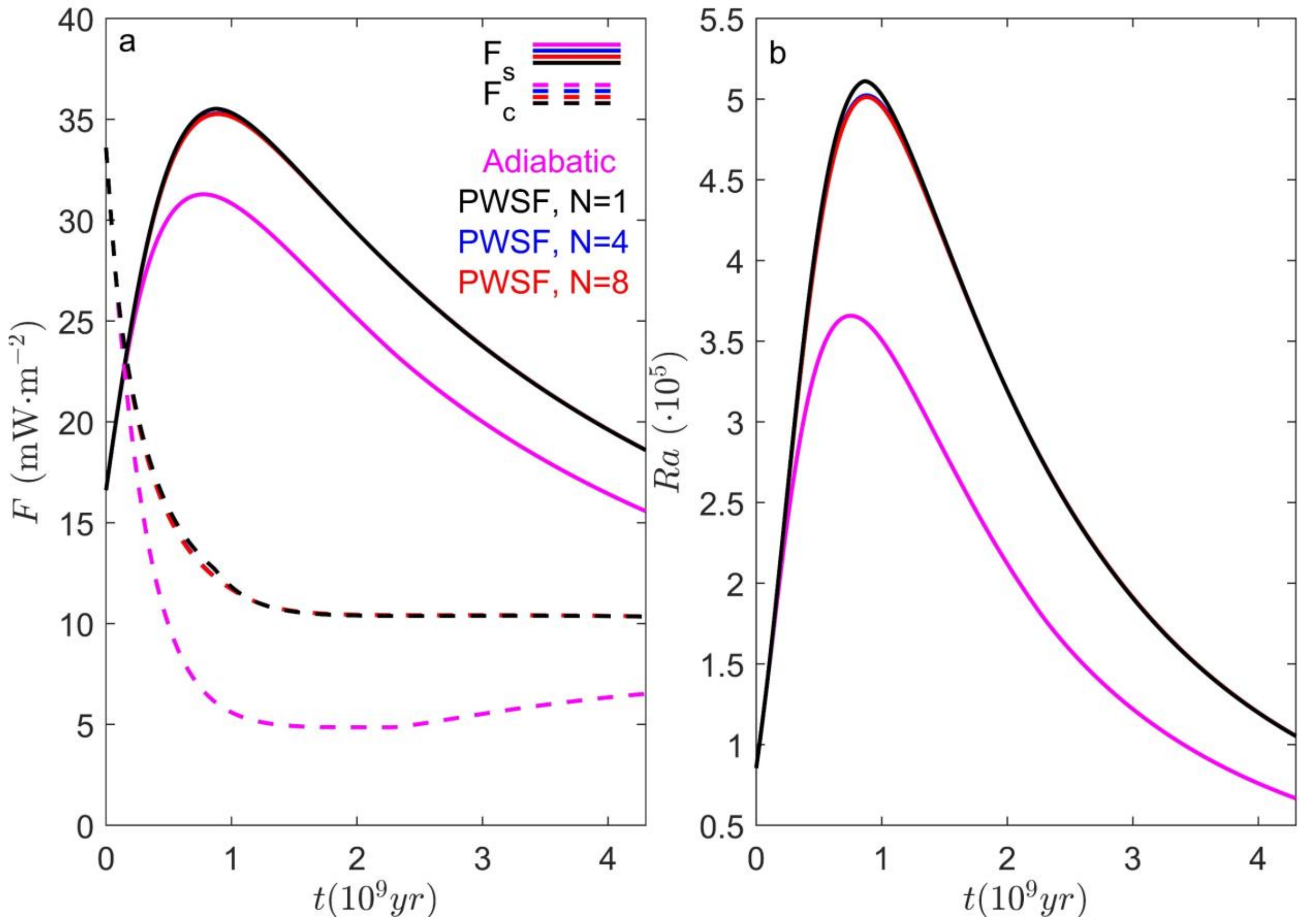

Caption of figure 6: Characteristics of the mantle for thermal evolution runs of Mercury that are also presented in figure 5 (see scenario Mercury[2] in table 1 for the adopted parameters). (a) The heat flux from the core to the mantle ($F_c$, dashed lines) and from the mantle to the lithosphere ($F_s$, solid lines), for the PWSF thermal evolution schemes with $N = 1$ (black), $N = 4$ (blue), and $N = 8$ (red), and for the thermal evolution run where the core's temperature profile is fixed by the adiabat and thermal stratification is neglected (magenta). (b) The Rayleigh number of the mantle scaled by a factor $10^5$. Blue lines are almost completely overplotted by red lines.

Figure A1

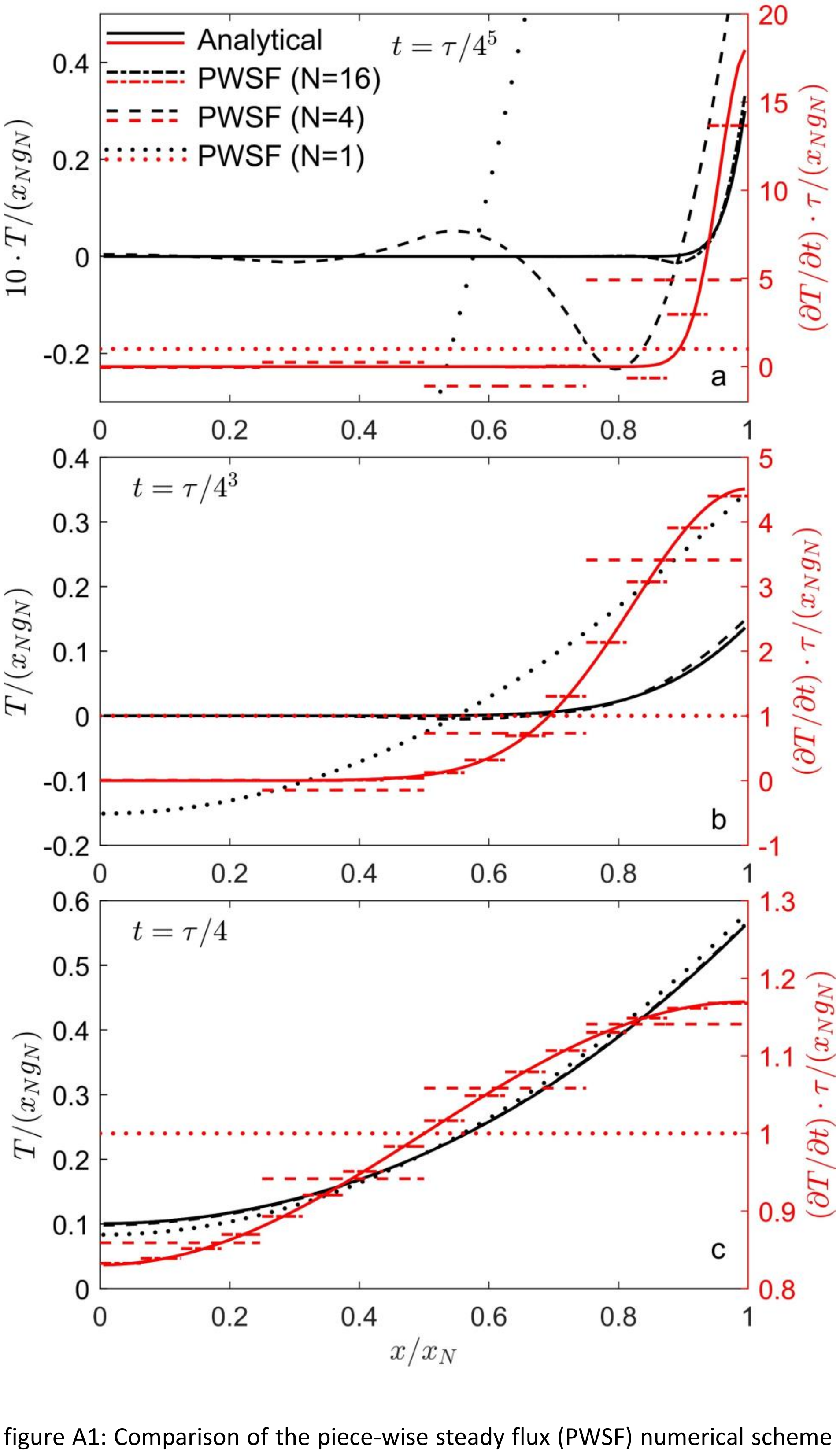

Caption of figure A1: Comparison of the piece-wise steady flux (PWSF) numerical scheme discretized into four intervals with $N = 16$ (dashed dotted), $N = 4$ (dashed) and $N = 1$ (dotted) with the analytical solution (solid) for conduction on a line segment $x_0 < x < x_N$, with $x_0 = 0$. Inhomogeneous

Neumann boundary conditions are adopted with gradient boundary conditions $\partial T(t, x_0)/\partial x = 0$, $\partial T(t, x_N)/\partial x = g_N$, and initial condition $T(0, x) = 0$. The diffusion timescale is $\tau = (x_N - x_0)^2/\kappa$, with $\kappa$ the diffusivity. Panels (a), (b), and (c) show the obtained normalized $T(t, x)/(x_N g_N)$ (black) at $t = \tau/4^5$, at $t = \tau/4^3$, $t = \tau/4$. Also, the normalized cooling rate $\big((\partial T/\partial t) \cdot \tau/(x_N g_N)\big)$ of the analytical solution is plotted (red solid) and compared to the non-dimensional secular cooling parameters of the temperature profile $(-S(r) \cdot \tau/(r_N g_N))$ of the PWSF numerical schemes, which are constant over each interval. The temperature of panel (a) is magnified relative to panels (b) and (c), to discriminate the temperature profile of the PWSF scheme with $N = 16$ from the analytical solution.

Figure C1

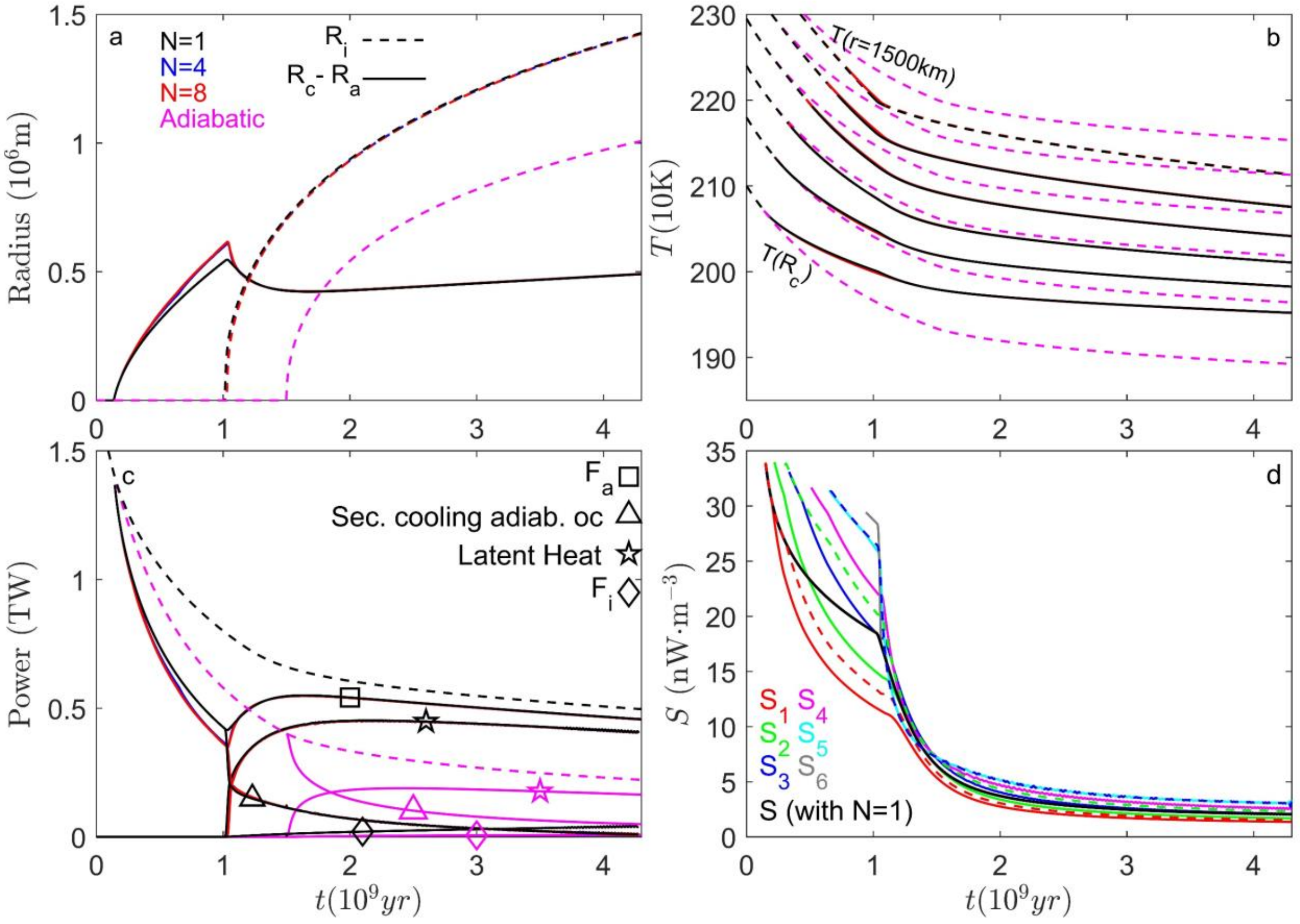

Caption of figure C1: Thermal evolution of Mercury's core, coupled to the thermal boundary layer of the mantle (see scenario Mercury[1] in table 1 for the adopted parameters). (a) The radius of the inner core ($R_i$) and the thickness of the conductive layer ($R_c - R_a$) set out against time. (b) Temperature at a radius of 1500 km, 1600 km, 1700 km, 1800 km, 1900 km radius, and at the core-mantle boundary ($R_c$) of 2020 km. The dashed and solid lines indicate the adiabatic and conductive temperature profile, respectively. (c) The energy balance of the adiabatic region of the liquid outer core. The heat that is transported across the interface ($R_a$) is denoted by $F_a$ (square symbol). The secular cooling in the adiabatic part of the liquid core is denoted by the triangle symbol. The release of latent (and gravitational) energy upon solidification is denoted by pentagram symbol. The heat flux through the inner core boundary, which equals the secular cooling of the solid inner core, is denoted by diamond symbol. (Panels a, b, and c) Thermal evolution runs with $N = 1$, $N = 4$, and $N = 8$ are plotted in black, blue and red, respectively. Blue lines are almost completely overplotted by red lines, whereas red lines are largely overplotted by black lines. Magenta lines correspond to a thermal evolution run where thermal stratification is not considered and the core's profile is considered to be adiabatic for the entire evolution. (d) The parameter $S$ in the thermal evolution with $N = 1$ is plotted in black. The $S_i$ for $i$ from 1 to 6, plotted by solid-coloured lines, are the parameters of the temperature profile across the six uppermost conductive intervals in the thermal evolution produced with $N = 8$. The $S_i$ for $i$ from 1 to 3, plotted by dashed-coloured lines, are the parameters of the temperature profile across the three uppermost conductive intervals in the thermal evolution produced with $N = 4$.

Table 1

| parameter | Symbol | Earth[1] | Earth[2] | Mercury[1] | Mercury[2] | Unit |
|---|---|---|---|---|---|---|
| Core density | $\rho$ | 10200 | 10200 | 7200 | 7200 | kg·m$^{-3}$ |
| Core heat capacity | $c$ | 860 | 860 | 840 | 840 | J·kg$^{-1}$·K$^{-1}$ |
| Core radius | $R_c$ | 3480 | 3480 | 2020 | 2020 | 10$^3$m |
| Planet radius | $R_p$ | - | - | 2440 | 2440 | 10$^3$m |
| Core thermal conductivity | $k$ | 60 | 60 | 40 | 40 | W·m$^{-1}$·K$^{-1}$ |
| Latent and gravitational heat | $L + E_g$ | - | - | 5 | 5 | 10$^5$J·K$^{-1}$ |
| Entropy change | $\delta_S$ | 118 | 118 | - | - | |
| Density jump at $R_i$ | $\Delta_\rho$ | 600 | 600 | - | - | kg·m$^{-3}$ |
| Adiabatic coefficient | $T_1$ | 0 | 0 | 0 | 0 | K·m$^{-1}$ |
| Adiabatic coefficient | $T_2$ | -2.8·10$^{-14}$ | -2.8·10$^{-14}$ | 0 | 0 | K·m$^{-2}$ |
| Adiabatic coefficient | $T_3$ | 1.692·10$^{-21}$ | 1.692·10$^{-21}$ | -2.2·10$^{-20}$ | -2.2·10$^{-20}$ | K·m$^{-3}$ |
| Initial temperature at $R_c$ | $T_c$ | 4247.3 | 4300 | 2100 | 2100 | K |
| Pressure at $R_c$ | $P_c$ | 135 | 135 | 5 | 5 | GPa |
| Mantle temperature | $T_m$ | - | 2630 | 1800 | 1800[a] | K |
| Mantle thermal expansion | $\alpha_m$ | - | 2 | 2 | 2 | 10$^{-5}$K$^{-1}$ |
| Mantle gravity | $g_m$ | - | 10 | 4 | 4 | m·s$^{-2}$ |
| Mantle density | $\rho_m$ | - | 4500 | 3200 | 3200 | kg·m$^{-3}$ |
| Mantle heat capacity | $c_m$ | - | 1200 | 1200 | 1200 | J·kg$^{-1}$·K$^{-1}$ |
| Mantle thermal conductivity | $k_m$ | - | 4 | 4 | 4 | W·m$^{-1}$·K$^{-1}$ |
| Viscosity parameter | $v_0$ | - | 4000 | 4000 | 4000 | m$^2$·s$^{-1}$ |
| Viscosity parameter | $A$ | - | 102000 | 50000 | 50000 | K |
| Critical Rayleigh number | $Ra_{cr}$ | - | 2000 | 2000 | 2000, 500[b] | - |
| Initial radiogenic heat production | $Q_0$ | - | - | - | 1.5·10$^{-11}$ | W·Kg$^{-1}$ |
| Half-life | $\tau_{1/2}$ | - | - | - | 1.9 | 10$^9$yr |
| Radius of the lithospheric base | $R_l$ | - | - | - | 2240 | 10$^3$m |

Caption of table 1: Parameter values for thermal evolution scenarios for Earth and for Mercury, which are described in detail in Appendix B. Thermal evolution scenario Earth[1] is a reproduction of Labrosse et al. (1997), with results presented in figure 3. In thermal evolution scenario Earth[2], Earth's core is coupled to a constant temperature mantle and the results are presented in figure 4. Mercury[1] is a preliminary thermal evolution scenario for Mercury with its core coupled to a constant temperature mantle, which results are presented in Appendix C. Scenario Mercury[2] is the thermal evolution scenario for Mercury with core coupled to a parametrized thermal behaviour of the mantle, which results are presented in figures 5 and 6. [a]In contrast to the other scenarios, $T_m$ varies with time in Mercury[2] and only the initial mantle temperature is listed in this table. [b] In thermal evolution scenario Mercury[2], a critical Rayleigh number of 500 is used to parametrize the upper thermal boundary layer of the convective mantle. The critical Rayleigh number of 2000 is used for the lower thermal boundary layer of the convective mantle.